\documentclass[aip,
               jcp,
               preprint,
               amsmath,
               superscriptaddress,
               floatfix,
               nobibnotes,
               ]{revtex4-2}

\usepackage{amssymb}
\usepackage{amsmath}
\usepackage{mathtools}
\usepackage{latexsym}
\usepackage{amsfonts}
\usepackage[mathscr]{eucal}
\usepackage{bm}
\usepackage[dvips]{graphicx}
\usepackage{color}
\usepackage[utf8]{inputenc}
\usepackage[flushleft]{caption}
\usepackage{subfigure}
\usepackage{rotating}
\usepackage{txfonts}
\usepackage{booktabs}
\usepackage[unicode]{hyperref}
\DeclareMathAlphabet{\mathpzc}{OT1}{pzc}{m}{it}

\def\figfoot{Chem-Moving-Mol}
\def\mathbi#1{\textbf{\em #1}}

\def\mi{\textrm{i}}

\makeatletter

\makeatletter

\newcommand{\figcaption}[2]{
    \noindent {\bf Figure \ref{#1}:} #2
    \vspace{1cm}}

\begin{document}


\title{The $[3+1]$ Formulation of Chemical Dynamics in Curved Spacetime
under the Eulerian Observer}

\author{Xingyu Zhang}
 \affiliation{Theoretical Chemistry,
              Department of Chemistry, 
              Northwestern Polytechnical University,
              West Youyi Road 127, 710072 Xi'an,
              China}

\author{Jinke Yu}
 \affiliation{Theoretical Chemistry,
              Department of Chemistry, 
              Northwestern Polytechnical University,
              West Youyi Road 127, 710072 Xi'an,
              China}
                            
\author{Qingyong Meng*}
\thanks{To whom correspondence should be addressed}
 \email{qingyong.meng@nwpu.edu.cn}
  \affiliation{Theoretical Chemistry,
               Department of Chemistry, 
               Northwestern Polytechnical University,
               West Youyi Road 127, 710072 Xi'an,
               China}

\date{\today}


\begin{abstract}

\noindent {\bf Abstract}:
Traditionally, gravity is generally considered to exert an extremely
weak effect in chemistry because the Newtonian gravitation is typically
negligible compared to the dominant Coulomb potentials in a molecular
system. In this work, we porpose a primitive framework of chemical
dynamics in curved spacetime through fiducial-observer $[3+1]$ formulation
by revising the nuclear Hamiltonian operator through the metric tensor
of configuration space rather than by adding Newtonian gravitation in
the potential energy term, where the absolute-sapce and universal-time
viewpoint of Galileo is adopted. Using frames fixed on normal observers
in the $[3+1]$ formalism ensures possibility of this treatment. Taking
spherically symmetric curved spacetime ({\it i.e.} Schwarzschild spacetime)
as numerical demonstration, we explore
(1) the H + H$_2$ reaction dynamics,
(2) the H$_2$ + H$_2$ scattering dynamics,
(3) dynamics of dissociative chemsorption of H$_2$O on Cu(111),
(4) the spectrum band of anthracene cation, and
(5) the Berry phase in the nuclear wave function of a 98D surface
scattering model. These calculations predict that (i) reaction or
scattering probability and (ii) spectrum band decrease abruptly to
zero as the spacetime curvature increases; meanwhile, the geometric
phase is unaffected by the spacetime curvature. Finally, discussions
on these numerical results, together with perspectives on the
applications of quantum field theory to chemical dynamics in curved
spacetime are given.
{\it Schwarzschild Spacetime}; {\it Nuclear Hamiltonian}

\end{abstract}
\maketitle

\section{Introduction\label{sec:intro}}

Our understanding on chemical reaction and molecular properties
\cite{yan20:582,xie20:767,che21:936,wan21:938,wan23:191,li24:746,rei25:962,zha25:20397,son24:597,mia24:532,son22:11128,men21:2702}
stems from empirical observations on earth with very weak gravity and
low velocit, called Newtonian limit or weak-gravitational-field
approximation. However, we inevitably encounter chemical phenomena
that demand new explanations when a molecular system livies in the
four-dimensional (4D) spacetime region near celestial bodies with
enormous mass or navigate in close proximity to them. These scenarios
compel us to further develop formalism theory of chemistry in curved
spacetime by extending the current theory in flat spacetime. Main goal
of this work is to primitively resolve such problems by developing a
theoretical scheme on chemical dynamics in a curved spacetime based on
the normal-congruence $[3+1]$ formulation \cite{sch12:book,mis17:book,car19:book}.
In this formulation, the $[3+1]$ Arnowitt-Deser-Misner (ADM) decomposition
\cite{sch12:book,mis17:book,car19:book} is adopted to split 4D spacetime
into a foliation of spacelike hypersurfaces $\Sigma_t$, labeled by a
temporal coordinate $t$, where the reference frames are fixed on the
normal congruence, also called the Eulerian observer. As well discussed
in general relativity 
\cite{tho82:339,sch12:book,nob16:032108,que16:102101,agg16:2653,mis17:book,car19:book},
when using these reference frames one can accept the concept the
adsolute-space and universal-time viewpoint of Galileo. Since current
chemistry adopts such viewpoint, it becomes possible to extend it in
flat spacetime into the present framework for curved spacetime.

In flat spacetime, solving the nuclear Schr{\"o}dinger equation (SE),
that is $\mathrm{i}\partial_t\Psi(t)=\hat{H}\Psi(t)$, one can directly
explore chemical dynamics \cite{zha25:20397,son24:597,mia24:532,son22:11128,men21:2702}
by either examining the transformation between the eigenstates of
reactant and product, known as time-independent quantum scattering,
or investigating the time evolution of the wavepacket, known as
time-dependent wavepacket propagation. Here, $\Psi(t)$ represents the
nuclear wave function and $\hat{H}=\hat{T}+\hat{V}$ the nuclear
Hamiltonian operator, where $\hat{T}$ and $\hat{V}$ operators are
kinetic energy operator (KEO) and potential energy surface (PES),
respectively. In these approaches, requirement of the flat spacetime
is frequently overlooked and supposed to be always true \cite{zha25:20397}.
This work, on the other hand, presents a theoretical framework of
chemical dynamics in curved spacetime based on the Eulerian-observer
$[3+1]$ decomposition
\cite{tho82:339,sch12:book,nob16:032108,que16:102101,agg16:2653,mis17:book,car19:book}
by revising the nuclear Hamiltonian operator through the metric of
configuration space, where the adsolute-space and universal-time
viewpoint of Galileo is adopted. To numerically demonstrate such
revisions, the H + H$_2$ \cite{son19:114302,son22:1983}, H$_2$ + H$_2$ \cite{son22:1983},
H$_2$O/Cu(111) \cite{son22:6047}, and photoelectron (PE) spectrum of
anthracene cation \cite{men13:014313} are employed through the
multiconfigurational time-dependent Hartree (MCTDH) product method and
its multilayer version (ML-MCTDH)
\cite{mey90:73,man92:3199,bec00:1,mey09:book,mey12:351,wan03:1289,man08:164116,ven11:044135,wan15:7951}.
For simple demonstrations, the spherically symmetric curved spacetime,
called the Schwarzschild spacetime is adopted, which is generated by a
massive object with no charge and rotation.

It is worth noting that, by directly adding the Newtonian gravitation
into the PES term, the SE has drawn increasing attentions to be often
employed as the conceptual and technical platforms in understanding the
interaction of quantum particle with weak gravity under the Newtonian
limit in flat spacetime \cite{ana14:085007,bas17:193002,ana22:015602,bos25:015003}.
In such Schr{\"o}dinger-type framework, spacetime curvature effects are
represented by the metric and quantum effects are reproduced by
non-relativistic quantum dynamics \cite{ana14:085007,bas17:193002,ana22:015602,bos25:015003},
which is the non-relativistic limit of quantum field theory (QFT).
However, when the limit of general relativity and non-relativistic
limit are taken on face, many theoretical points are ignored. Recently,
a perspective and discussion on extending the Schr{\"o}dinger theory
of quantum molecular dynamics in flat spacetime to a QFT-based
formulation was made \cite{zha25:20397} by extending single-particle
approximation to field theory. Based on the present theory and previous
perspective \cite{zha25:20397}, in this work we will further discuss
the perspectives of chemical dynamics in curved spacetime beyond the
present theoretical framework.

The rest of this paper is organized as follows. Section \ref{sec:theory}
describes details of the theoretical framework of this work and numerical
details of the benchmarks. Section \ref{sec:sm-chem} gives the numerical
results and discussions. Finally, Section \ref{sec:con} concludes with
a summary. Unless otherwise specified, the natural units are always used,
where the physical constants are given by $\hbar=\epsilon_0=c=G=1$.
Moreover, by Einstein summation convention, repeated upper and lower
indices always imply summation in this work.

\section{Theoretical Framework\label{sec:theory}}

\subsection{$[3+1]$ Decomposetion under the Eulerian Observer\label{sec:assum}}

Given in Table \ref{tab:assum-curved} are assumptions of the present
work. First of all, the present thoery requires $[3+1]$ decomposition
\cite{sch12:book,mis17:book,car19:book} of the 4D spacetime which is
globally foliated into a one-parameter family of spacelike three-dimensional
(3D) slices $\Sigma_t$ and a time coordinate $t$ (see Figure
\ref{fig:reax-spacetime}), where the metric tensors for 4D spacetime
and 3D slice are denoted by $g_{\mu\nu}$ and $\gamma_{ij}$, respectively.
To illustrate the present framework simply, the equations of motion
(EOMs) of the spacetime and the molecular system are assumed to be
completely decoupled, while the geometric characteristics of $\Sigma_t$
are taken to be time-invariant. Thus, the present theoretical framework
predicts results from the viewpoint of the Eulerian observer whose
worldline is orthogonal to 3D slice and denoted by $n^{\mu}$ as
illustrated in Figure \ref{fig:reax-spacetime}. Based on the Eulerian
observer, one can accept the concept the adsolute-space and universal-time
viewpoint of Galileo
\cite{tho82:339,sch12:book,nob16:032108,que16:102101,agg16:2653,mis17:book,car19:book}.
By the above assumption, incorporating curved spacetime into quantum
molecular dynamics in this work is straightforward by revising the
nuclear Hamiltonian operator. Nevertheless, problem which may start to
arise must be noted is arisen from the PES term, which is often already
constructed through electronic energies computed by electronic-structure
calculations. In this work, the PES term is supposed to be constructed
in flat spacetime because the orders-of-magnitude difference between
electron and nucleus ($\sim10^{-15}\;\mathrm{m}$ versus $\gtrsim10^{-12}\;\mathrm{m}$)
allows neglect of curvature effects on the electronic motion itself.
Moreover, the Coulomb interaction in a uniformly curved spacetime
\cite{nob16:032108,que16:102101,agg16:2653} with a constant positive
curvature $K$ is revised in the form
$V^{(\mathrm{curved})}_{\mathrm{coul}}=
V^{(\mathrm{flat})}_{\mathrm{coul}}\sqrt{1-Kr_{\mathrm{e}}^2}$,
where $V^{(\mathrm{flat})}_{\mathrm{coul}}$ is the Coulomb interaction
in flat spacetime. The curvature radius, $\sqrt{1/K}$, can be seen as
apparent radius of the universe of order of magnitude $10^{26}\;\mathrm{m}$,
while $r_{\mathrm{e}}$ is approximately equal to atomic radius of $10^{-10}\;\mathrm{m}$.
Therefore, it is safety to have approximation $V^{(\mathrm{curved})}_{\mathrm{coul}}\approx
V^{(\mathrm{flat})}_{\mathrm{coul}}$ for electrons moving in the molecular
system. In addition, due to the acceptance of absolute time, problem arisen
from relativity effects no longer exists in this work. However, it is
worth noting that, when one includes additional relativity effects, the
inner product between quantum states defined in the case with the existence
of absolute time is not Lorentz invariant \cite{bos25:015003}. If one
wants to describe chemical dynamics in a manner consistent with relativity,
a covariant formalism must be adopted to make the EOMs and the inner
products Lorentz invariant \cite{bos25:015003,zha25:20397}. The QFT in
either flat or curved spacetime enables it in the low-energy regime. We
will return to discussions on this point later.

\begin{figure}[h!] 
 \subfigure[\quad Mapping from curved spacetime to configuration space]{%
  \includegraphics[width=18cm]{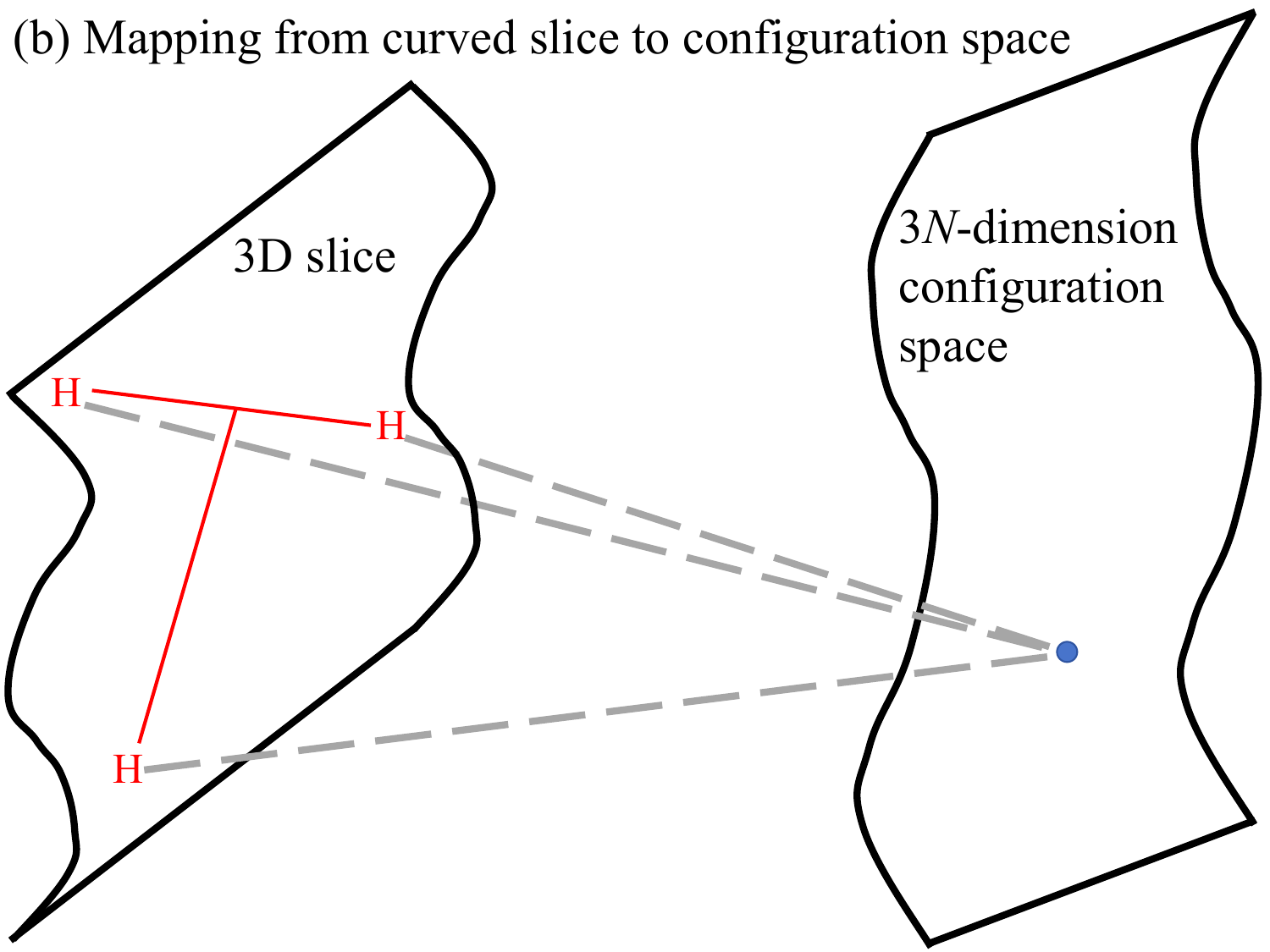}}
   \caption{\figfoot}
    \label{fig:map-slice-conf}
     \end{figure}

Turning to molecular representation, as illustrated by Figure \ref{fig:map-slice-conf},
the conformation of an $N$-atomic molecule in 3D slice can be
homeomorphically represented by a single point in a $3N$-dimensional
smooth manifold, called configuration space, locally homeomorphic to
$\mathbb{R}^{3N}$. Such a homeomorphism mapping between 3D slice and
configuration space exists because they share certainly topological
invariants. The geometry of configuration space is similarly
characterized by the metric $\tilde{\gamma}^{ij}$ for the Cartesian
coordinates. By a set of generalized coordinates $\mathbi{q}$, the KEO
has been thoroughly derived through the polyspherical approach
\cite{gat09:1,ndo12:034107,ndo13:204107,sad12:234112} in the form
\begin{equation}
2\hat{T}=-J^{-1}\partial_{\iota}J\varg^{\iota\kappa}\partial_{\kappa},\quad
J=\left\vert\det\left(\frac{\partial x^{(i)}}{\partial q^{(\kappa)}}
\right)\right\vert=\Big\vert\det\Big(\partial_{\kappa}x^{(i)}\Big)\Big\vert,
\label{eq:polysph-keo-011}
\end{equation}
where the metric $\varg^{\iota\kappa}$ for $\mathbi{q}$ is given by
$\varg^{\iota\kappa}=\tilde{\gamma}^{ij}\partial_{i}q^{(\iota)}
\partial_{j}q^{(\kappa)}$ and $\varg_{\iota\kappa}=(\varg^{\iota\kappa})^{-1}$
leading to $J=(\vert\det(\varg^{\iota\kappa})\vert)^{-1/2}$. 
For simple, the Schwarzschild metric is chosen in this work as a
benchmark, where the spacetime is around a static and spherically
symmetric gravitating source with mass $M$ in an uncharged and
non-rotating vacuum. The event interval is then given by
\begin{align}
\mathrm{d}s^2&=-\left(1-\frac{2M}{r}\right)\mathrm{d}t^2+
\left(1-\frac{2M}{r}\right)^{-1}\mathrm{d}r^2
+r^2\mathrm{d}\vartheta^2+r^2\sin^2\vartheta\mathrm{d}\varphi^2
\allowdisplaybreaks[4] \nonumber \\
&=-\Big(1-\varrho\Big)\mathrm{d}t^2+
\Big(1-\varrho\Big)^{-1}\mathrm{d}r^2
+r^2\mathrm{d}\vartheta^2+r^2\sin^2\vartheta\mathrm{d}\varphi^2,
\label{eq:schd-metric-000-revi}
\end{align}
where $r_s$ is the Schwarzschild radius $r_s$ and $r$ the distance
between the system and the mass source $M$. The dimensionless parameter
$\varrho=2M/r=r_s/r$ determines the strength of influence of gravity.
In Table \ref{tab:num-varrho-values}, we give values of $\varrho$
used for the present benchmarks and give physical insight of these
$\varrho$ values.

\begin{figure}[h!]
   \centering
    \subfigure[\quad Bath part of ML-tree of the 96D + 2D wave function]{%
     \includegraphics[width=23cm,angle=90]{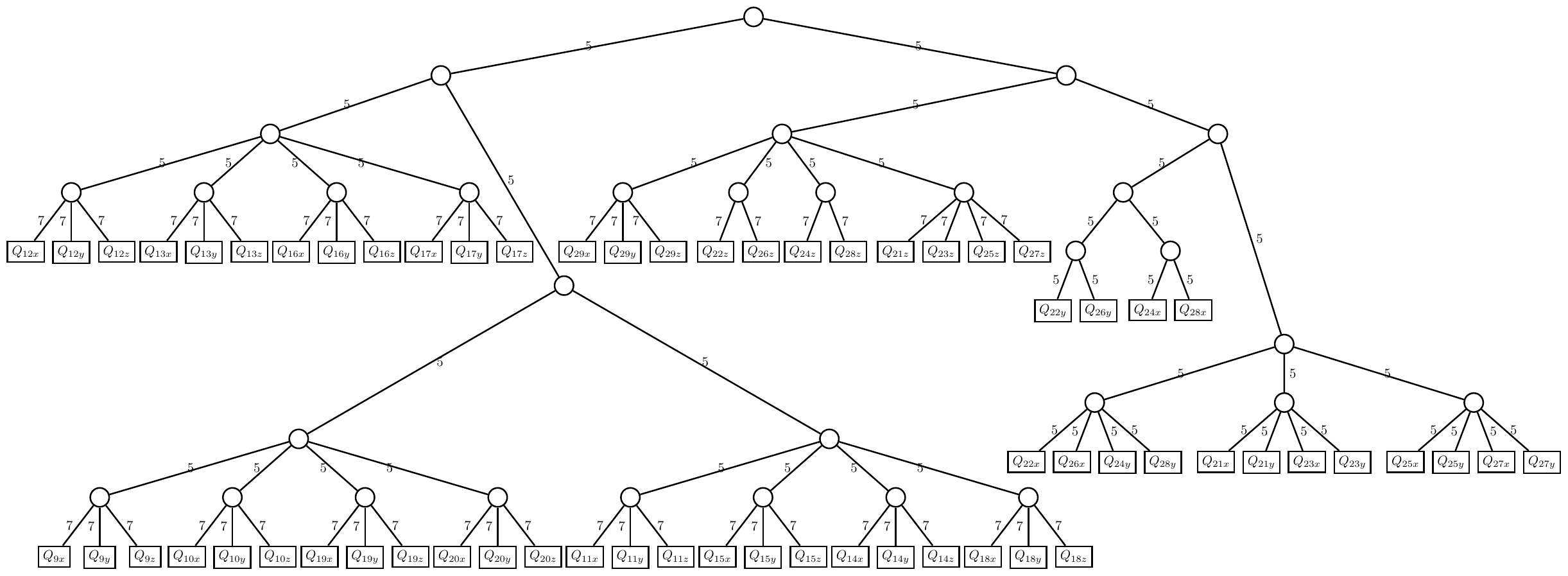}}
     \caption{\figfoot}
      \label{fig:ml-tree-98d}
       \end{figure}

In this work, we take several kinds of systems as benchmarks,
including (1) the H + H$_2$ reaction, (2) the H$_2$ + H$_2$ scattering,
(3) dissociative chemsorption of H$_2$O on Cu(111), (4) the spectrum band
of the anthracene cation, and (5) the geometric phase in the nuclear wave
function of a model for CO/Cu(100). These systems represent typical
processes in chemical dynamics \cite{men22:16415,zha25:20397}. According
to Equations \eqref{eq:polysph-keo-011} and \eqref{eq:schd-metric-000-revi}
together with transform of coordinate frames, one can revise the KEOs in
the flat spacetime obtaining those in the Schwarzschild spacetime. To this
end, the molecular systems are positioned at a distance $r$ from the
origin point and oriented the reaction or scattering
coordinate along the radial direction. Moreover, the entire systems
are supposed not to undergo orbital rotation around the spacetime
origin. This is possible because, as given above, we restrict the
treatment to the classical regime, where absolute time still exists
and thus the molecular velocity must remain relatively low. By adding
the KEO and PES, we obtain nuclear Hamiltonian operator for subsequent
ML-MCTDH calculations. In ML-MCTDH, the total time-dependent nuclear
wave function is expressed in terms of a tensor in a hierarchical Tucker
format, represented by a tree-like structure (called ML-tree), where
the basis contains products of single-particle functions (SPFs). Figure
\ref{fig:ml-tree-98d} illustrates the optimal ML-trees for the present
ML-MCTDH calculations. Inserting
the multilayer ansatz into the Dirac-Frenkel variational principle, the
ML-MCTDH working equations for arbitrary layering schemes and for all
intermediate quantities have been well derived
\cite{wan03:1289,man08:164116,ven11:044135,wan15:7951}. These coupled
non-linear differential equations can be efficiently solved using
standard numerical tools \cite{bec00:1,wan03:1289,man08:164116,ven11:044135}.
In computing reaction or scattering probabilities, we analyse the flux
of the wave function fraction through the separation surface by adding
into the Hamiltonian operator a complex absorbing potential (CAP)
\cite{ris93:4503,ris96:1409},
$V_{\mathrm{CAP}}=-\mathrm{i}\eta\cdot(s-s^{(\mathrm{CAP})})^n\cdot
\Theta(s-s^{(\mathrm{CAP})})\cdot\Theta(t-t^{(\mathrm{CAP})})$,
where $n$ and $\eta$ are order and strength of the CAP, while $s$ and
$s^{(\mathrm{CAP})}$ represent reaction coordinate and the separation
surface, respectively. Here, $\Theta(\cdot)$ is step function and
$t^{(\mathrm{CAP})}$ means the time that the CAP appears. By flux
analysis through the CAP, we can then compute the probability as function
of collision energy $E$. For the spectrum band of anthracene cation,
assuming the Condon approximation, the Fourier transform of autocorrelation
function $C(t)=\langle\Psi(0)\vert\Psi(t)\rangle$ predicts spectrum band
\cite{men13:014313}. Here, we also use the so-called $t/2$-trick for the
autocorrelation function, which doubles the length of the autocorrelation
function. We refer the reader to the first SI file for further details of
the present quantum dynamics calculations.

\subsection{Models for Reaction and Spectrum\label{sec:curved-revision}}

As shown in Section \ref{sec:assum} and Table \ref{tab:assum-curved},
the Schwarzschild (denoted by ``schd'' here) metric is chosen for
deriving the KEOs of the present benchmarks. A simple way to do this
is to directly modify the KEOs in the flat spacetime to obtain those
in the Schwarzschild spacetime. We refer the reader to the first
Supporting Information (SI) file for details of the KEOs in the flat
spacetime for the present benchmarks. For simple calculations, setting
coordinates $\mathbi{q}$ to be spherical polar coordinates $\{r,\vartheta,
\varphi\}$, the Schwarzschild metric defined by Equation
\eqref{eq:schd-metric-000-revi} satisfies
\begin{equation}
g_{\mu\nu}^{(\mathrm{schd})}=\left(\begin{array}{cccc}
-(1-2M/r) & 0 & 0 & 0 \\
0 & (1-2M/r)^{-1} & 0 & 0  \\
0 & 0 & r^2 & 0 \\
0 & 0 & 0 & r^2\sin^2\vartheta \\
\end{array}\right)=
\left(\begin{array}{cccc}
-(1-\varrho) & 0 & 0 & 0 \\
0 & (1-\varrho)^{-1} & 0 & 0  \\
0 & 0 & r^2 & 0 \\
0 & 0 & 0 & r^2\sin^2\vartheta \\
\end{array}\right).
\label{eq:schd-metric-keo-001}
\end{equation}
In essence, the ratio $\varrho=2M/r=r_s/r$ is a fundamental dimensionless
parameter which quantifies the strength of gravity in a spherically
symmetric spacetime without any rotation and charge. Given in Table
\ref{tab:num-varrho-values} and the first SI file are physical regime
associated with each range of $\varrho$. It should be noted that, as
illustrated by Figure \ref{fig:map-slice-conf}, the coordinates to
define the spectime metric and those to define configuration space of
the molecular system must be distinguished. The former (used to obtain
Equation \eqref{eq:schd-metric-keo-001}) are spherical polar coordinates,
while the latter must be specifically designed for configuration space.
In this work, due to their reactivity, the H + H$_2$, H$_2$ + H$_2$,
and H$_2$O/Cu(111) systems are represented by Jacobi-type coordinates
$\{r_{\mathrm d},r_{\mathrm v},\theta\}$, $\{b,r_1,r_2,\theta_1,\phi_1,
\theta_2,\phi_2\}$, and $\{x,y,z\}\oplus\{r_1,r_2,u_1,\theta_1,\theta_2,
\phi\}$, respectively. Moreover, for nonreactive systems, the anthracene
cation and model of CO/Cu(100), the normal coordinates (or called
normal modes) are adopted. We refer the reader to the first SI file for
the definitions of these coordinates and conformations of the benchmarks,
together with numerical setup of the present MCTDH calculations for
the models given in this subsection.

For the H + H$_2$ system, according to the KEO in the flat space, the
KEO in the Schwarzschild sapcetime can be revised as
\begin{align}
\hat{T}_3^{(\mathrm{schd})}&=-\left(1-\frac{2M}{r}\right)
\frac{1}{2\mu_{\mathrm{H-H}_2}}\frac{\partial^2}{\partial r^2_{\mathrm d}}
-\frac{1}{2\mu_{\mathrm{H}_2}}\frac{\partial^2}{\partial r^2_{\mathrm v}}
-\left[\left(1-\frac{2M}{r}\right)\frac{1}{2\mu_{\mathrm{H-H}_2}r^2_{\mathrm d}}+
\frac{1}{2\mu_{\mathrm{H}_2}r^2_{\mathrm v}}\right]
\left(\frac{1}{\sin\theta}\frac{\partial}{\partial\theta}\sin\theta
\frac{\partial}{\partial\theta}\right)
\allowdisplaybreaks[4]  \nonumber \\
&=-\frac{1-\varrho}{2\mu_{\mathrm{H-H}_2}}\frac{\partial^2}{\partial r^2_{\mathrm d}}
-\frac{1}{2\mu_{\mathrm{H}_2}}\frac{\partial^2}{\partial r^2_{\mathrm v}}
-\left[\frac{1-\varrho}{2\mu_{\mathrm{H-H}_2}r^2_{\mathrm d}}+
\frac{1}{2\mu_{\mathrm{H}_2}r^2_{\mathrm v}}\right]
\left(\frac{1}{\sin\theta}\frac{\partial}{\partial\theta}\sin\theta
\frac{\partial}{\partial\theta}\right).
\label{eq:triatom-ham-curved-999}
\end{align}
For the H$_2$ + H$_2$ system, the KEO in the Schwarzschild spacetime
is similarly given by
\begin{align}
\hat{T}_4^{(\mathrm{schd})}={}&-\frac{1-\varrho}{2\mu}
\frac{\partial^2}{\partial b^2}-\sum_{i=1}^2\frac{1}{2\mu_i}
\frac{\partial^2}{\partial r_i^2}
-\left(\frac{1-\varrho}{2\mu b^2}+\frac{1}{2\mu_1r_1^2}\right)
\left[\frac{1}{\sin\theta_1}\frac{\partial}{\partial\theta_1}
\sin\theta_1\frac{\partial}{\partial\theta_1}
+\frac{1}{\sin^2\theta_1}\frac{\partial^2}{\partial\phi_1^2}\right]
\allowdisplaybreaks[4] \nonumber \\
&-\left(\frac{1-\varrho}{2\mu b^2}+\frac{1}{2\mu_2r_2^2}\right)\left[
\frac{1}{\sin\theta_2}\frac{\partial}{\partial\theta_2}
\sin\theta_2\frac{\partial}{\partial\theta_2}
+\frac{1}{\sin^2\theta_2}\frac{\partial^2}{\partial\phi_2^2}\right]
+\frac{1-\varrho}{2\mu b^2}\left[2\frac{\partial^2}{\partial\phi_1^2}\right.
\allowdisplaybreaks[4]  \nonumber \\
&+\exp\big(+\mi\phi_1\big)\left(+\frac{\partial}{\partial\theta_1}
+\mi\cot\theta_1\frac{\partial}{\partial\phi_1}\right)\times
\exp(-\mi\phi_2)\left(-\frac{\partial}{\partial\theta_2}
+\mi\cot\theta_2\frac{\partial}{\partial\phi_2}\right) 
\allowdisplaybreaks[4]  \nonumber \\
&+\left.\exp\big(-\mi\phi_1\big)\left(-\frac{\partial}{\partial\theta_1}
+\mi\cot\theta_1\frac{\partial}{\partial\phi_1}\right)\times\exp(+\mi\phi_2)
\left(+\frac{\partial}{\partial\theta_2}
+\mi\cot\theta_2\frac{\partial}{\partial\phi_2}\right)\right] 
\allowdisplaybreaks[4]  \nonumber \\
&+\frac{1-\varrho}{2\mu b^2}\left[J(J+1)+2\frac{\partial}{\partial\phi_2}
\left(\frac{\partial}{\partial\phi_1}+\frac{\partial}{\partial\phi_2}\right)\right] 
\allowdisplaybreaks[4]  \nonumber \\
&-(1-\varrho)\frac{C_+(J,K)}{2\mu b^2}\left[\exp\big(+\mi\phi_1\big)
\left(+\frac{\partial}{\partial\theta_1}
+\mi\cot\theta_1\frac{\partial}{\partial\phi_1}\right)+\exp(+\mi\phi_2)
\left(+\frac{\partial}{\partial\theta_2}
+\mi\cot\theta_2\frac{\partial}{\partial\phi_2}\right)\right] 
\allowdisplaybreaks[4]  \nonumber \\
&-(1-\varrho)\frac{C_-(J,K)}{2\mu b^2}\left[\exp\big(-\mi\phi_1\big)
\left(-\frac{\partial}{\partial\theta_1}
+\mi\cot\theta_1\frac{\partial}{\partial\phi_1}\right)
+\exp(-\mi\phi_2)\left(-\frac{\partial}{\partial\theta_2}
+\mi\cot\theta_2\frac{\partial}{\partial\phi_2}\right)\right],
\label{eq:tetraatom-ham-curved-999}
\end{align}
where $C_{\pm}=\sqrt{J(J+1)-K(K\pm1)}$. For both Equations \eqref{eq:triatom-ham-curved-999}
and \eqref{eq:tetraatom-ham-curved-999}, the values of $\varrho$ are
given in Table \ref{tab:num-varrho-values}, while the total angular
momentum is set to be zero, that is $C_{\pm}=0$. Furhtermore, the KEO
of the H$_2$O/Cu(111) system in the Schwarzschild spacetime can be
given by
\begin{align}
\hat{T}_{\mathrm{wat}}^{(\mathrm{schd})}={}&-\frac{2}{3m_{\mathrm{wat}}}
\left(\frac{\partial^2}{\partial x^2}+
\frac{\partial^2}{\partial y^2}+\frac{\partial^2}{\partial x\partial y}\right)
-\frac{1-\varrho}{2m_{\mathrm{wat}}}\frac{\partial^2}{\partial z^2}
-\frac{1}{2\mu_{\mathrm{OH}}}\frac{\partial^2}{\partial r_1^2}
-\frac{1-\varrho}{2\mu_{\mathrm{H-OH}}}\frac{\partial^2}{\partial r_2^2}
\allowdisplaybreaks[4]  \nonumber \\
&-\frac{1}{2}\left(\frac{1}{\mu_{\mathrm{OH}}r_1^2}+
\frac{1-\varrho}{\mu_{\mathrm{H-OH}}r_2^2}\right)
\frac{\partial}{\partial u_1}\big(1-u_1^2\big)\frac{\partial}{\partial u_1}
\allowdisplaybreaks[4]   \nonumber \\
&+\frac{1}{2}\left(\frac{1}{\mu_{\mathrm{OH}}r_1^2}+
\frac{1-\varrho}{\mu_{\mathrm{H-OH}}r_2^2}\right)
\frac{\hat{J}_{z^{\mathrm{BF}}}^2}{1-u_1^2}+
\frac{1-\varrho}{2\mu_{\mathrm{H-OH}}r_2^2}
\big(\hat{J}^2-2\hat{J}^2_{z^{\mathrm{BF}}}\big)
\allowdisplaybreaks[4] \nonumber \\
&-\frac{1-\varrho}{4\mu_{\mathrm{H-OH}}r_2^2}
\left(\frac{\partial}{\partial u_1}\sqrt{1-u_1^2} 
+\sqrt{1-u_1^2}\frac{\partial}{\partial u_1}\right)
\big(\hat{J}_{+^{\mathrm{BF}}}-\hat{J}_{-^{\mathrm{BF}}}\big)
\allowdisplaybreaks[4]   \nonumber \\
&+\frac{1-\varrho}{4\mu_{\mathrm{H-OH}}r_2^2}\frac{u_1}{\sqrt{1-u_1^2}}
\Big[\big(\hat{J}_{+^{\mathrm{BF}}}+\hat{J}_{-^{\mathrm{BF}}}\big)\hat{J}_{z^{\mathrm{BF}}}+
\hat{J}_{z^{\mathrm{BF}}}\big(\hat{J}_{+^{\mathrm{BF}}}+\hat{J}_{-^{\mathrm{BF}}}\big)\Big].
\label{eq:keo-4externel-curved-sche}
\end{align}
To obtain Equation \eqref{eq:keo-4externel-curved-sche}, as given in the
end of Section \ref{sec:assum}, the water molecule is positioned at a
distance from the spacetime origin and approximately oriented the reaction
coordinate along the radial direction of the Schwarzschild spacetime.
Again, values of $\varrho$ in Equation \eqref{eq:keo-4externel-curved-sche}
are given in Table \ref{tab:num-varrho-values}.

For the anthracene cation, we adopt the full-dimensional multistate
multimode vibronic coupling Hamiltonian (MMVCH) model 
\cite{koe84:59,gha11:14523,red10:111102,men13:014313}. The diabatic
electronic representation is employed with six electronic states, where
the diagonal elements are adiabatic PESs of these electronic states
while the off-diagonal coupling matrix elements are derived from the
vibronic couplings. Thus, the Hamiltonian matrix is given by $6\times6$
matrix
\begin{equation}
\mathbf{H}_{6\times6}^{(\mathrm{flat})}=
\Big(\hat{T}_{\mathrm{an}}^{(\mathrm{flat})}+
\hat{V}_{\mathrm{an}}^{(\mathrm{flat},0)}\Big)\mathbf{1}_{6\times6}+
\mathbf{W}_{6\times6}^{(\mathrm{flat})}=\frac{1}{2}\sum_{i=1}^{66}\omega_i
\left(-\frac{\partial^2}{\partial v_i^2}+v_i^2\right)
\mathbf{1}_{6\times6}+\mathbf{W}_{6\times6}^{(\mathrm{flat})},
\label{eq:ana-ham-matrix-000}
\end{equation}
where $\omega_i$ is the vibrational frequency of the $i$-th normal
mode. The operator $\hat{T}_{\mathrm{an}}^{(\mathrm{flat})}+
\hat{V}_{\mathrm{an}}^{(\mathrm{flat},0)}$
is the summation of the vibrational energies of all normal modes (see
the second SI file for definitions), denoted by $\{v_i\}_{i=1}^{66}$.
The elements of the $\mathbf{W}_{6\times6}^{(\mathrm{flat})}$ matrix
are expanded as a Taylor series in $v_i$ around the Frank-Condon point,
where linear plus quadratic terms for the diagonal elements and linear
terms for the off-diagonal elements are retained \cite{gha11:14523,red10:111102,men13:014313}.
All of the parameters involved in the present MMVCH model were reported
in References \cite{gha11:14523}. Since the vibrational normal modes
are linear coordinates of the Cartesian coordinates, one should consider
the spectrum band of the anthracene cation in Cartesian system. However,
it is difficult to derive the Hamiltonian matrix by re-writting Equation
\eqref{eq:ana-ham-matrix-000} in Cartesian system. Physically, this is
because the Schwarzschild spacetime is spherically symmetric rather
than plane symmetry for the spectrum of the molecular system. As given
in Section \ref{sec:assum}, we adopt the simplest way to consider the
spectrum in the curved spacetime, by which the anthracene molecule is
far from the origin of the Schwarzschild spacetime. Thus, the Hamiltonian
operator in the Schwarzschild spacetime is given by
\begin{equation}
\mathbf{H}_{6\times6}^{(\mathrm{schd})}=
\Big(\hat{T}_{\mathrm{an}}^{(\mathrm{schd})}+
\hat{V}_{\mathrm{an}}^{(\mathrm{schd},0)}\Big)\mathbf{1}_{6\times6}+
\mathbf{W}_{6\times6}^{(\mathrm{schd})}=\left[\frac{1}{2}
\sum_{i=1}^{66}\omega_i\left(-\Big(1-\varrho\Big)
\frac{\partial^2}{\partial v_i^2}+\Big(1-\varrho\Big)^{-1}v_i^2\right)\right]
\mathbf{1}_{6\times6}+\mathbf{W}_{6\times6}^{(\mathrm{schd})},
\label{eq:plane-wave-ana-spec-900}
\end{equation}
where the potential matrix $\mathbf{W}_{6\times6}^{(\mathrm{schd})}$
satisfies
\begin{align}
W_{YY}^{(\mathrm{schd})}&=W_{YY}^{(\mathrm{flat})}=E_Y+\sum_{i\in{A_g}}
\kappa_i^{(Y)}v_i+\frac{1}{2}
\sum_{\mathrm{all~modes}}\gamma_i^{(Y)}
v_i^2,
\label{eq:plane-wave-ana-spec-901} \allowdisplaybreaks[4] \\
W_{YZ}^{(\mathrm{schd})}&=W_{YZ}^{(\mathrm{flat})}=\sum_{i\in{\Gamma_Y\otimes\Gamma_Z}}
\lambda_i^{(Y,Z)}v_i.
\label{eq:plane-wave-ana-spec-902}
\end{align}
In both Equations \eqref{eq:plane-wave-ana-spec-901} and \eqref{eq:plane-wave-ana-spec-902},
we set $Y$ and $Z$ to be in the set $\{\tilde{X},\tilde{A},\tilde{B},
\tilde{C},\tilde{D},\tilde{E}\}$. Again, the $\varrho$ values are given
in Table \ref{tab:num-varrho-values}. In addition, another practical
way adopts the metric of the plane gravitational wave. However, at the
present its use is still theoretically impossible due to exsitence of
the retarded time and the inequality between the temporal and spatial
coordinates in the molecular EOMs. Later, a perspective on this issue
and a way to resolve this problem are given. 

\subsection{Model for Molecular Geometric Phase in Surface Scattering\label{sec:nuclear-phase}}

Now, let us turn to numerical demonstrations on high-dimensional model
for time-dependent phase in curved spacetime. The present numerical
model describes a diatomic molecule CO adsorbed on movable Cu(100)
surface. The Cu(100) surface is supposed to be a three-layer grid with
$30$ atoms with mass $m_3=63.546$ dalton and coordinates $\{Q_{i\mu}
\}_{i=0}^{29}$, where the atom \#0 is the top site, \#1-\#8 the first
layer, \#9-\#20 the second layer, \#21-\#29 the third layer. The
molecular motions are abstractly described by coordinates set
$\{b,\chi,\phi\}\oplus\{R,\vartheta,z\}$. The internal coordinates set
$\{b,\chi,\phi\}$ describes the C-O stretch and two-dimensional rotation
of CO with polar angle $\chi$ and azimuthal angle $\phi$, while $\{R,
\vartheta,z\}$ describes rotation of the entire molecule along a loop
with radius $R_0$ and angle $\vartheta$. Thus, there are a total of
$96$ dimensionalities. We refer the reader to the first SI file for
conformation of this surface scattering model. The Hamiltonian operator
can be written as summation of the KEO for the slow modes
$\hat{T}_{\mathrm{slow}}$ and the Hamiltonian operator for the fast
modes $\hat{H}_{\mathrm{fast}}'$,
\begin{align}
\hat{H}(\mathbi{z})&=\hat{T}_{\mathrm{slow}}(\mathbi{Q},\boldsymbol{\lambda})
+\hat{H}_{\mathrm{fast}}'(\mathbi{z})=
\hat{T}_{\mathrm{slow}}(\mathbi{Q},\boldsymbol{\lambda})+
\hat{T}_{\mathrm{fast}}(\mathbi{q},\boldsymbol{\lambda})+
V_{\mathrm{fast}}(\mathbi{q},\mathbi{Q}_0,\boldsymbol{\lambda})+
V_{\mathrm{slow}}(\mathbi{Q},\boldsymbol{\lambda})
+V_{\mathrm{coul}}(\mathbi{z})
\allowdisplaybreaks[4] \nonumber \\
&=\hat{H}_{\mathrm{mol}}+\hat{H}_{\mathrm{surf}}'+\hat{H}_{\mathrm{3D-CHO}}=
\hat{T}_{\mathrm{mol}}+\hat{V}_{\mathrm{mol}}+\hat{T}_{\mathrm{surf}}'
+\hat{V}_{\mathrm{surf}}'+\hat{H}_{\mathrm{3D-CHO}},
\label{eq:model-ham-000}
\end{align}
where $V_{\mathrm{fast}}$ is the PES of the fast modes with the slow
modes fixed at $\mathbi{Q}_0$, $V_{\mathrm{slow}}$ the PES for the
other modes, and $V_{\mathrm{coul}}$ the coupling term. Moreover, we
let $\hat{H}_{\mathrm{mol}}$ and $\hat{H}_{\mathrm{surf}}'$ describe
motions of CO and Cu(100), respectively.

Moreover, the 3D coupled harmonic
oscillator (CHO) model \cite{men15:164310,men17:184305,men21:2702,men22:16415}
is used to give the coupling term $\hat{H}_{\mathrm{3D-CHO}}$ between
CO and Cu(100) \cite{men15:164310,men17:184305,men21:2702,men22:16415}
\begin{align}
2\hat{H}_{\mathrm{3D-CHO}}\big(q_1,q_2,q_3,\eta,\xi\big)&=
\frac{\hat{p}_1^2}{m_1}+\frac{\hat{p}_2^2}{m_2}+\frac{\hat{p}_3^2}{m_3}+
m_3\omega_3^2q_3^2+m_1\omega_1^2(q_1-q_2)^2+m_2\omega_2^2(q_2-q_3)^2
\allowdisplaybreaks[4] \nonumber  \\
&=\hat{\tilde{p}}_1^2+\hat{\tilde{p}}_2^2+\hat{\tilde{p}}_3^2+
\omega_3^2\tilde{q}_3^2+\omega_1^2\left(\tilde{q}_1-\zeta_1\tilde{q}_2\right)^2
+\omega_2^2\left(\tilde{q}_2-\zeta_2\tilde{q}_3\right)^2
\allowdisplaybreaks[4] \nonumber  \\
&=\hat{\tilde{p}}_1^2+\hat{\tilde{p}}_2^2+\hat{\tilde{p}}_3^2+\omega_2^2
\left(\begin{array}{ccc}
\tilde{q}_1 & \tilde{q}_2 & \tilde{q}_3
\end{array} \right)
\left(\begin{array}{ccc}
\eta         & -\zeta_1\eta    & 0  \\
-\zeta_1\eta & 1+\zeta_1^2\eta & -\zeta_2  \\
0            & -\zeta_2        & \xi+\zeta_2^2
\end{array}\right)
\left(\begin{array}{c}
\tilde{q}_1  \\
\tilde{q}_2  \\
\tilde{q}_3
\end{array}\right),
\label{eq:3d-cho-hamiltonian-00}
\end{align}
where $\tilde{q}_i=\sqrt{m_i}q_i$ and $\tilde{p}_i=p_i/\sqrt{m_i}$ with
$i=1,2,3$, are mass-scaled coordinates and momenta, respectively, while
$\zeta_1=\sqrt{m_1/m_2}$ and $\zeta_2=\sqrt{m_2/m_3}$ are coupling parameters.
It is worth noting that, parameters $\omega_1$ and $\omega_3$ are frequencies
of the molecule and the top surface atom, respectively, and thus parameters
$\eta=\omega_1^2/\omega_2^2$ and $\xi=\omega_3^2/\omega_2^2$ represent
two additional conditions of the present 96D surface scattering. The
parameters $\eta$ and $\xi$ introduce the nozzle and surface temperatures,
respectively. Therefore, the present 96D + 2D dynamics calculations
represent the 96D surface scattering process in varying 2D parameter
space $\{\eta,\xi\}$. The geometric phase is arisen from the curvature
of the parameter space. Next, the Hamiltonian $\hat{H}_{\mathrm{mol}}$
for CO reads
\begin{align}
\hat{H}_{\mathrm{mol}}={}&\hat{T}_{\mathrm{mol}}+\hat{V}_{\mathrm{mol}}
\allowdisplaybreaks [4] \nonumber \\
={}&-\frac{1}{2m}\left[\frac{1}{b^2}\frac{\partial}{\partial b}
\left(b^2\frac{\partial}{\partial b}\right)+
\frac{1}{b^2\sin\chi}\frac{\partial}{\partial\chi}
\left(\sin\chi\frac{\partial}{\partial\chi}\right)+
\frac{1}{b^2\sin^2\chi}\frac{\partial^2}{\partial\phi^2}\right]
\allowdisplaybreaks [4] \nonumber \\
&-\frac{1}{2(m_1+m_2)}\left(\frac{1}{R}\frac{\partial}{\partial R}
R\frac{\partial}{\partial R}+
\frac{1}{R^2}\frac{\partial^2}{\partial\vartheta^2}\right)+
\frac{1}{2}k_b\big(b-b_0\big)^2+\frac{1}{2}k_R\big(R-R_0\big)^2,
\label{eq:model-ham-001}
\end{align}
where $m=m_1m_2/(m_1+m_2)$ is the resuced mass of CO. The radial potential
$\hat{V}_{\mathrm{mol}}=k_R(R-R_0)^2/2$ ensures that the CO molecule always
move along the closed loop. Motions of the Cu(100) surface are described
by \cite{men21:2702}
\begin{align}
\hat{H}_{\mathrm{surf}}'=-\frac{1}{2m_3}\sum_{\mu=x,y}
\frac{\partial^2}{\partial Q_{0\mu}^2}-
\frac{1}{2m_3}\sum_{i=1}^{29}\sum_{\mu=x,y,z}
\frac{\partial^2}{\partial Q_{i\mu}^2}
+\sum_{i=0}^{29}\sum_{\mu=x,y,z}
V_{\mathrm{M}}\big(Q_{i\mu}-Q_{i\mu}^{(0)}\big)+
\frac{1}{2}\sum_{0\leq i<j\leq29}k_Z\Big(Q_{iz}-Q_{jz}\Big)^2,
\label{eq:model-ham-002}
\end{align}
where $Q_{i\mu}^{(0)}$ represents optimized geometry of the $i$-th surface
atom and $V_{\mathrm{M}}(\rho)$ is a Morse potential with parameters $D$
and $a$,
\begin{equation}
V_{\mathrm{M}}(\rho)=D\big[\exp(-2a\rho)-2\exp(-a\rho)\big]
\label{eq:model-ham-003}
\end{equation}
Such surface model has been used to study surface scattering dynamics
of the CO/Cu(100) system \cite{men21:2702}. One should note that the
3D CHO Hamiltonian $\hat{H}_{\mathrm{3D-CHO}}$ already contains the
Hamiltonian for out-of-plane motion of the top atom and the coupling
term between molecule and surface.

With the above Equations \eqref{eq:3d-cho-hamiltonian-00}, \eqref{eq:model-ham-001},
and \eqref{eq:model-ham-002}, one can revise the KEO of the present 96D
+ 2D model in the flat space by introducing the parameter $\varrho$
appeared in the Schwarzschild metric (see Equations \eqref{eq:schd-metric-000-revi}
and \eqref{eq:schd-metric-keo-001}). For simple, as indicated in the
end of Section \ref{sec:assum} we still assume that the distance between
molecule and surface is the radial cooridante in the Schwarzschild
spacetime. Thus, Equation \eqref{eq:3d-cho-hamiltonian-00} is rewritten
as
\begin{align}
2\hat{H}_{\mathrm{3D-CHO}}^{(\mathrm{schd})}\big(q_1,q_2,q_3,\eta,\xi\big)&=
\frac{\hat{p}_1^2}{m_1}+\frac{\hat{p}_2^2}{m_2}+\frac{(1-\varrho)\hat{p}_3^2}{m_3}+
\frac{m_3\omega_3^2q_3^2}{1-\varrho}+
m_1\omega_1^2(q_1-q_2)^2+m_2\omega_2^2\left(q_2-\frac{q_3}{\sqrt{1-\varrho}}\right)^2
\allowdisplaybreaks[4] \nonumber  \\
&=\hat{\tilde{p}}_1^2+\hat{\tilde{p}}_2^2+(1-\varrho)\hat{\tilde{p}}_3^2+
\frac{\omega_3^2\tilde{q}_3^2}{1-\varrho}+\omega_1^2\left(\tilde{q}_1-\zeta_1\tilde{q}_2\right)^2
+\omega_2^2\left(\tilde{q}_2-\frac{\zeta_2\tilde{q}_3}{\sqrt{1-\varrho}}\right)^2.
\label{eq:3d-cho-hamiltonian-99900}
\end{align}
Table \ref{tab:98d-model} gives parameters of the present 96D CHO + 2D
Hamiltonian operator. We refer the reader to the first SI file for
numerical details of the present 96D + 2D ML-MCTDH calculations and
the multilayer tree-structure (ML-tree) of the 96D + 2D wave function.

\section{Results and Discussions\label{sec:sm-chem}}

\subsection{Dynamics Results\label{sec:relativity}}

\begin{figure}[h!]
 \centering
  \includegraphics[width=18cm]{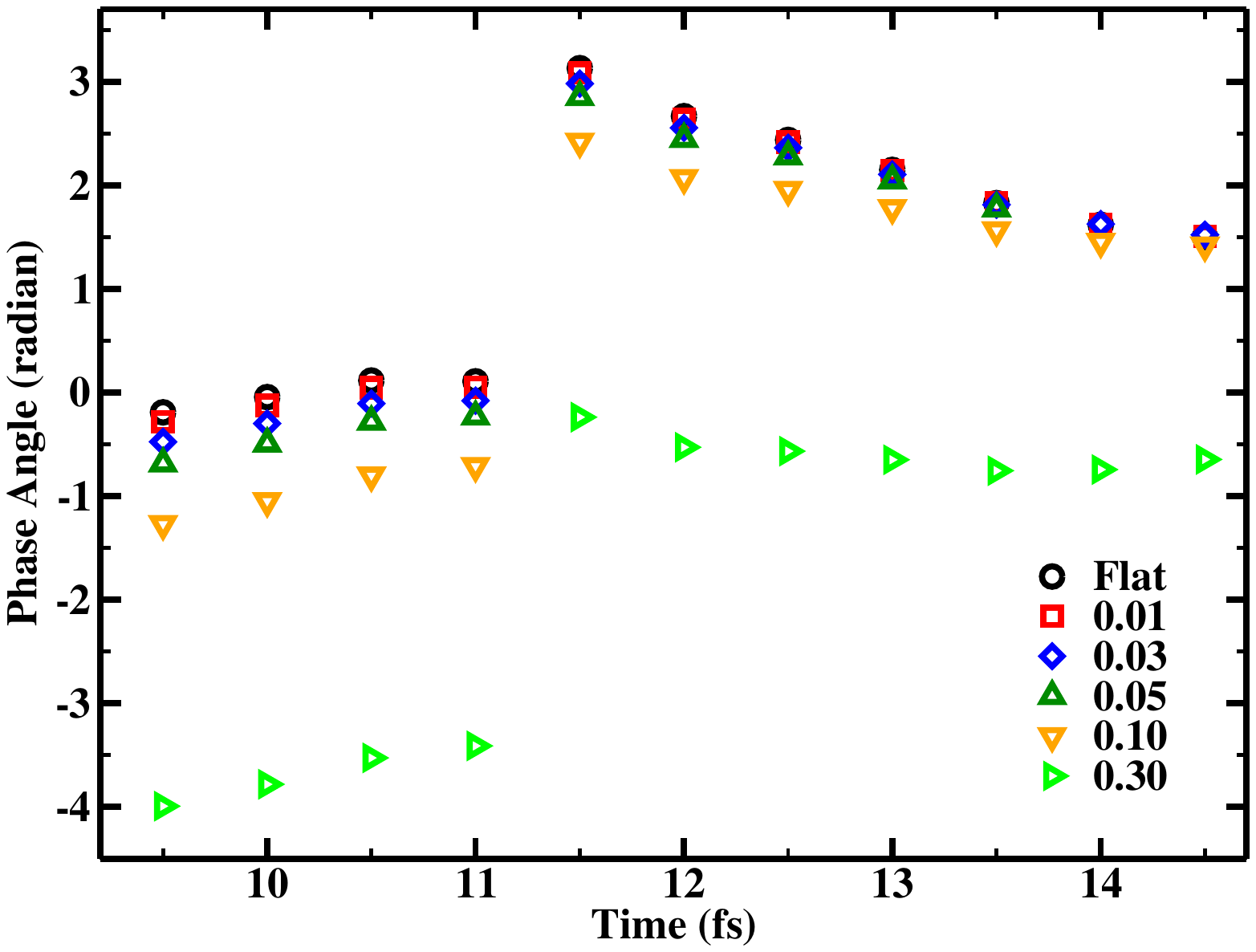}
   \caption{\figfoot}
    \label{fig:berry-phase-var}
     \end{figure}

Illustrated by Figure \ref{fig:num-results} are comparisons of dynamics
results in the Schwarzschild sapcetime with those in flat spacetime for H +
H$_2$, H$_2$ + H$_2$, H$_2$O/Cu(111), and anthracene cation in
$\tilde{B}\;{}^2A_u$. Figure \ref{fig:berry-phase-var} illustrates
relation between $\varrho$ and the time-dependent geometric phase of
the nuclear wave function. In these figures, the black lines are results
in the flat space. The cyan, maroon, light yellow, light green, violet,
yellow, green, blue, and red lines represent reactive or scattering
probabilities for $\varrho$ values of $0.60$, $0.50$, $0.40$, $0.30$,
$0.20$, $0.10$, $0.05$, $0.03$, and $0.01$, respectively. The present
numerical results demonstrate the following two point. First, the
spacetime curvature (and hence the gravitational field) drastically
reduces the reaction probability, scattering probability, spectrum
band making these molecular properties approach to zero at
$\varrho>0.60$. Second, at an appropriate strength, the
spacetime curvature appears to amplify the lowest-energy resonance
of the reaction if there exists one. Third, the spacetime curvature
causes the spectrum band to undergo a blueshift. Finally, unlike the
above cases, while the phase itself depends on $\varrho$, the transition
of the Berry phase is hardly influenced by the spacetime curvature, as
shown in Figure \ref{fig:berry-phase-var}. This might be arisen from
the fact that the Berry phase originates from the curvature of the
parameter space, which is generally independent of the spacetime
background. As a result, the spacetime curvature hardly influences the
phase transition.

For the H + H$_2$ reaction, {\it i.e.} typical chemical reaction in the
gas phase, Figure \ref{fig:num-results}(a) indicates that in strong gravitational
field with $\varrho>0.10$, all reaction probabilities exhibit a minor
peak in the low-energy region around $0.05$ eV and a large peak in the
high-energy region around $0.4\sim0.5$ eV. In particular, as $\varrho$
increases from $0.20$ to $0.60$, the reactive probability at the region
of $0.4\sim0.5$ eV increases and then decreases, reaching its maximum
of near unit at approximately $\varrho=0.30\sim0.40$. In this case, the
reactive probability drops abruptly to zero once the energy exceeds a
certain energy in the range from $0.6$ to $1.5$ eV, that increases rapidly
as $\varrho$ decreases. This is unexpected because the reactive probability
tends to approach a non-zero value once the impacting energy exceeds a
certain value for typical reactions. Furthermore, this situation changes
significantly when $\varrho$ falls below $0.10$, and the reaction
probability gradually approaches normal when $\varrho$ decreases below
$0.05$. An interesting phenomenon in the low-energy region of $0.3\sim0.6$
eV is that the reaction probability initially increases rapidly to
nearly $0.95$ as $\varrho$ increases from $0.00$ to $0.40$, then
approaches sharply to zero as $\varrho$ continues to increase from
$0.40$ to $0.60$. When $\varrho$ exceeds $0.60$, the reaction probability
remains almost invariably zero. For typical chemical reaction on the
surface, Figure \ref{fig:num-results}(c) illustrates dissociative
chemsorption probabilities of H$_2$O $\to$ H + OH on the Cu(111) surface.
In flat space, as previously discussed \cite{son22:6047} the dissociation
probability is approximately equal to zero, when the impacting energy
is less than roughly $0.15$ eV. At impacting energy of more than $0.15$
eV, the dissociation probability increase from $0$ to $\sim10^{-2}$.
In curved spacetime, the situation is similar to the results of H + H$_2$,
as shown in Figure \ref{fig:num-results}(a), where the dissociation probabilities
of higher-energy region become samller than that in flat space and
fast approach to nearly zero when $\varrho>0.10$. At low-energy region,
the situation becomes complex likely caused by numerical calculations.

Having given reaction probabilities of typical reactions in the gas
phase and on the surface, now we should turn to examples of molecular
properties. For the H$_2$ + H$_2$ scattering, as illustrated by Figure
\ref{fig:num-results}(b), the situation is similar to the cases of H + H$_2$
and H$_2$O/Cu(111), that is the scattering probability depends on the
value of $\varrho$. As the impacting energy increases, a larger value
of $\varrho$ leads to a rapid drop in the scattering probability from
unit to nearly zero. Noting that no reaction occurs for the H$_2$ +
H$_2$ scattering process, the scattering probability must be close to
unit in the flat space. However, the scattering probabilities with non-zero
values of $\varrho$ as shown in Figure \ref{fig:num-results}(b) imply that
the two H$_2$ molecules form a weak dimer at non-zero impacting energy.
This dimer mat be caused by the existence of gravitational field.
Furthermore, increasing the value of $\varrho$ accelerates the decay
rate of the scattering probability. For instance, the scattering
probabilities approach to zero at $\sim0.28$, $\sim0.45$, and $\sim1.5$
eV for the $\varrho$ values of $0.50$, $0.30$, and $0.10$, respectively.
Obviously, this is because of the existence of the curved space. In
this context, the enhanced gravity in the curved space may cause two
objects to mutually attract and form a dimer. Next, Figure \ref{fig:num-results}(d)
illustrates spectrum bands of the anthracene cation in the
$\tilde{B}\;{}^2A_u$ state at the cases with various values of $\varrho$.
The parameter $\varrho$ exerts a dual influence on the spectrum bands.
On one hand, the spectral intensity decreases with increasing $\varrho$
from zero. When $\varrho<0.10$, the change in spectral intensity remains
relatively insignificant. However, it becomes markedly evident once
$\varrho$ exceeds $0.10$. When $\varrho$ is large enough, the spectral
intensity tends to fast approach zero. On the other hand, the band center
shifts toward higher energy regions as $\varrho$ increases. When
$\varrho<0.10$, the spectral position shows minimal variation, with the
band center shifting slightly toward higher energy regions by approximately
$0.01$ eV (about $806$ cm$^{-1}$). In contrast, when $\varrho>0.10$,
the band shifts rapidly to higher energy regions. For instance, the
band positions are $9.21$, $9.31$, $9.52$, and $9.78$ eV at the
$\varrho$ values of $0.10$, $0.30$, $0.50$, and $0.60$, respectively.

Finally, by the present 96D + 2D ML-MCTDH calculations, the time-dependent
phase angle are shown in Figure \ref{fig:berry-phase-var}. The present
96D + 2D ML-MCTDH calculations indicate a transition of the phase angle
at $\sim11.5$ fs. Meanwhile, the time evolution of the phase exhibits
distinct nonlinear characteristics when the duration is either less
than or greater than $\sim11.5$ fs. This nonlinear feature indicates
ongoing energy transformations during the reaction process. At the
critical point, the phase angle changes by $\pi$ meaning the wave
function changes sign during the propagation in the parameter space
$\{\eta,\xi\}$. The expectation $\langle\xi\rangle$ of $1.976$ at
$\sim11.5$ fs is similar to the critical point at $1.119$. Further,
Figure \ref{fig:berry-phase-var} also indicates that the phase angles
at different $\varrho$ always change by $\pi$ at the critical point.
This implies that the geometric phase is hardly influenced by the
values of $\varrho$. It is worth considering reasons for such phenomenon.
As was discussed in detail
\cite{yar96:985,yar96:10456,yar98:971,yar01:6277,jua07:044317,bou08:124322,jua08:211101,alt08:214117,alt08:1,bou10:969,jan13:144316,gao15:12036,xia10:1959,ced04:3,bae06:boa,ced13:224110,yua18:1289,yan20:582,xie20:767,che21:936,wan21:938,pan22:545,wan23:191,li24:746,rei25:962},
the geometric phase effects on the electronic properties depend on the
geometric characteristics (say Berry curvature and connection) of the
parameter space for electronic motions, that is the nuclear configuration
space. Similarly, the geometric phase effects on nuclear properties
depend on the geometry of the parameter space for nuclear motions, such
as the space $\{\eta,\xi\}$. In differential geometry, the concept of
connection is introduced to compare vectors in nearby vector spaces
enabling us to find whether the direction of a vector is changed when
it parallel transports along a path. The parallel transport does not
change length and direction of the tangent vector of a geodesic, which
can be seen as a ``straight line'' in curved manifolds. The adiabatic
motion of the quantum state in flat Hilbert space is a typical example
of the parallel transport. In this case, the motion is unaffected by
the variation of external parameters. If the parameter space becomes
curved, the direction of a vector is changed when it parallel
transports along a close path. Keeping this idea in the mind, since
the parameter space is independent of the nuclear configuration space,
the curvature and connection of the latter does not affect the geometric
properties of the former. Therefore, we can obtain the present 96D + 2D
results at various values of $\varrho$ because these results only depend
on geometric characteristics of the parameter space rather than those
of the nuclear configuration space.

\begin{figure}[h!]
 \centering
  \subfigure[\quad Expectation of total energy of H$_2$O on Cu(111) in curved spacetime]{
   \includegraphics[width=18cm]{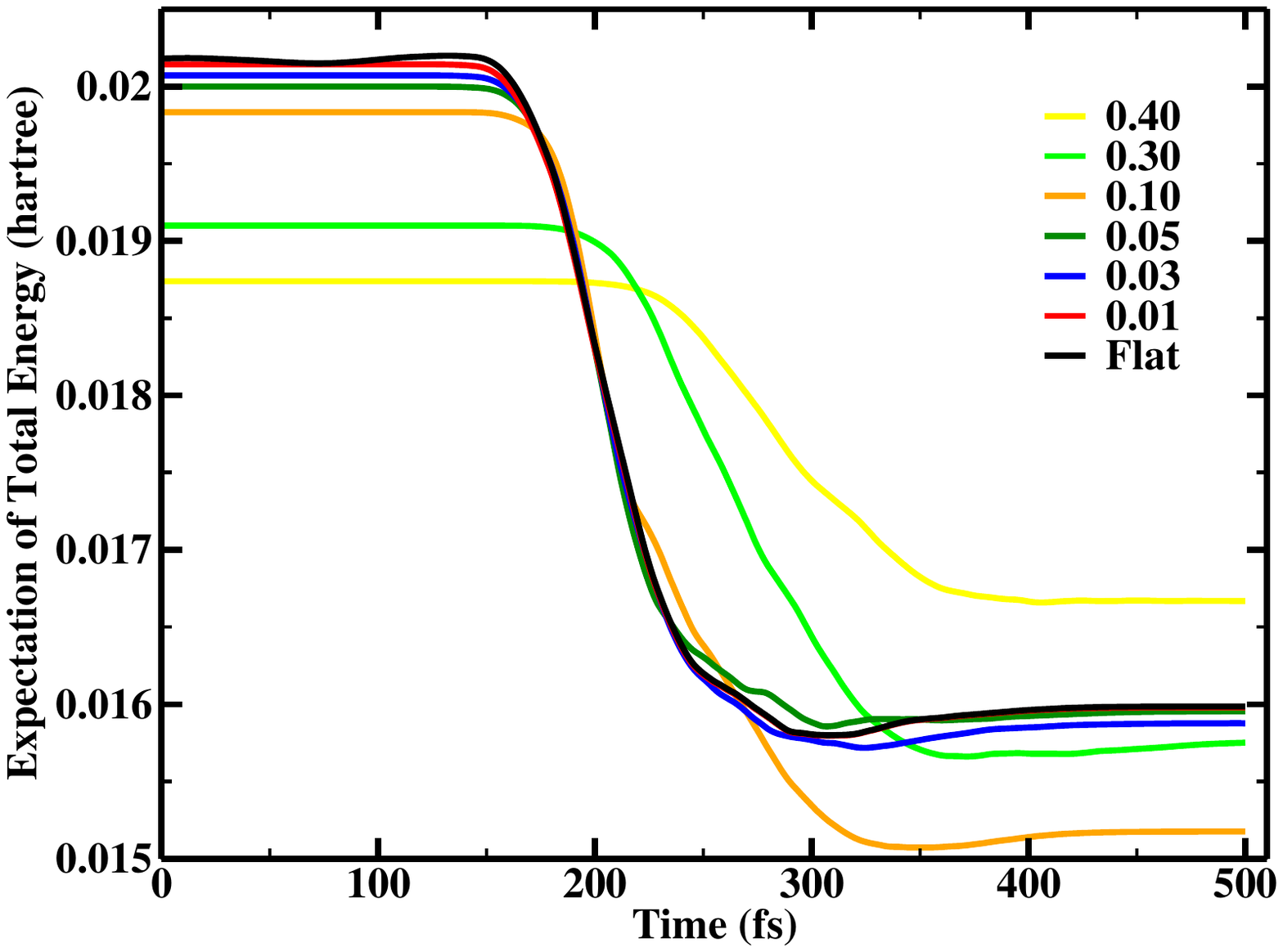}}
   \caption{\figfoot}
    \label{fig:exp-proper}
     \end{figure}

Having shown numerical results on dynamics in curved spacetime, we must
turn to analysis of the wave function and the time-dependent expectations
computed for the present benchmarks. Figure \ref{fig:norm-wavefunc}
illustrates comparsions of the square of the wave function norm
$\Vert\Psi(t)\Vert^2=\langle\Psi(t)\vert\Psi(t)\rangle$
at different values of $\varrho$, together with the autocorrection
$C(t)=\langle\Psi(0)\vert\Psi(t)\rangle$ of the anthracene cation.
The colored lines represent the results computed by various $\varrho$
values. Figure \ref{fig:exp-proper} illustrates comparsions of the
time-dependent expectation values of the total energy at different
values of $\varrho$. These expectations are computed by propagated
wave function $\Psi(t)$ through the expression
$\langle E\rangle(t)=\langle\Psi(t)\vert\hat{H}\vert\Psi(t)\rangle/
\langle\Psi(t)\vert\Psi(t)\rangle$,
where $\hat{H}$ is the Hamiltonian operator at different cases. The
colored lines represent the results computed by various $\varrho$
values. Here, we must emphasize that, since a CAP is employed in the
propagations for these systems, the wave function only satisfies
conserved normalization $\langle\Psi(t)\vert\Psi(t)\rangle=\mathrm{constant}$
in the early stage of propagation. Once the CAP becomes active, as
illustrated by Figure \ref{fig:norm-wavefunc}, conserved normalization
is lost, and $\langle E\rangle$ no longer satisfy conservation. For
this reason, we truncate the wave function norm and energy expectations
at some given time. As shown in Section \ref{sec:assum}, the present
theory adopts concept of absolute time and thus no special-relativity
effect is included in this work. Therefore, Figures \ref{fig:exp-proper}
and \ref{fig:norm-wavefunc} clearly indicate conservation of the total
energy. 

\subsection{Advantages and Disadvantages\label{sec:results-diss}}

Now, discussions on the present results shown in Section \ref{sec:relativity}
should be given to give advantages and disadvantages of the present
theory, as summarized in Tables \ref{tab:discuss-perspe} and
\ref{tab:discuss-future-theory}. First of all, numerical results predict
that the reaction probabilities and spectrum bands in the Schwarzschild
spacetime clearly depend on $\varrho$. This is not surprising because
$\varrho$ has been introduced in the KEO that plays one of importance
roles in quantum dynamics calculations (with the other importance
component being PES). As given by Equation \eqref{eq:schd-metric-000-revi},
value of $\varrho$ characterizes strength of the gravitational field
and related with effective mass of the reaction or scattering coordinate,
(called dynamical coordinate), which is consistent with the fact that
curvature of spacetime is associated with the mass. In this work, the
molecular systems are set to be positioned at a large enough distance
from the origin of Schwarzschild spacetime and be oriented the dynamical
coordinate along its radial direction. The former assumption effectively
localizes the entire molecule within an approximately confined region
of space, implying that each atom experiences a nearly uniform curvature
of the space. It allows for a consistent definition of the metric tensor
associated with individual atoms. The latter assumption implies that the
total rotations of the system as well as non-reactive motions remain unaffected
by the gravitational field. Therefore, one can ignore the angular elements
of the Schwarzschild metric and thus only revise the kinetics energy
terms associated with the dynamical coordinate. Moreover, the entire
systems is supposed to not undergo orbital rotation around the origin
of the spacetime making speed of the entire system remain much smaller
than the speed of light. It allows us to use the concept of absolute
time.

Second, the fact that the Schwarzschild spacetime is around a static
and spherically symmetric gravitating source with mass $M$ in an uncharged
and non-rotating vacuum precludes the emergence of electromagnetic
corrections \cite{nob16:032108,que16:102101,agg16:2653} and complex
rotation-vibration coupling (say Coriolis coupling) in the molecular
system which is a typical electromagnetic confinement system. This
allows us to easily obtain the KEO and PES terms of the nuclear
Hamiltonian. The other fact that the Schwarzschild spacetime is static
clearly allows for complete decoupling between the spacetime geometry
and the system. Nevertheless, the spacetime geometry, that is metric
and curvature, satisfies the ADM equations which might be coupled with
the EOM of the molecular system. Either existence or expression of such
coupled EOMs is still an open question. Phenomenologically, this problem
might be solved by deriving new Hamiltonian-formulation EOMs which
contain the Schr{\"o}dinger-type equation and the ADM equations. These
coupled EOMs may be derivable using a method analogous to that employed
in deriving the EOMs for relativistic hydrodynamics \cite{sch12:book,mis17:book,car19:book}.
A direct method might be to recognize that the molecular dynamics can
be seen as hydrodynamics in phase space, thereby allowing the direct
application of relativistic hydrodynamics \cite{sch12:book,mis17:book,car19:book}.
However, because we should abandon the viewpoint of Galileo (see Section
\ref{sec:theory}) such phenomenological EOMs remain unsatisfactory meaning
that we require spacetime-covariant EOMs. To overcome this problem, a
complete theory must incorporate spacetime- and/or Lorentz-covariant
EOMs in chemical dynamics, where the language of QFT might be required,
say those presented in Reference \cite{zha25:20397}. This is a fundamentally
change of the framework of current molecular quantum dynamics in flat
spacetime. Numerous previously reported attempts \cite{zha25:20397}
have demonstrated the feasibility of such change. In addition, extension
of QFT to curved spacetime should be further made to consider quantum
processes in gravitational field \cite{bir82:book,bar09:book,hac16:book}.
This is still not a full theory of quantum gravity.

Third, since the spherically symmetric Schwarzschild spacetime is neither
the sole solution to the Einstein field equations nor can satisfy general
spacetime, let us turn to other kind of curved spacetime. We provide some
typical solutions of the Einstein field equations and their physical
insight in the first SI file. As confirmed through experimental observation
\cite{tho18:040503}, the gravitational plane wave represents another
distinct type of solution to the Einstein field equations. The infinitesimal
interval in this spacetime with gravitational plane wave which propagates
along the $z$ axis satisfies
\begin{equation}
\mathrm{d}s^2=-\mathrm{d}t^2+\mathrm{d}z^2+(1+h_+(u))\mathrm{d}x^2+
(1-h_+(u))\mathrm{d}y^2+2h_{\times}(u)\mathrm{d}x\mathrm{d}y,
\label{eq:plane-grav-wave-metric}
\end{equation}
where $u=t-z$ is retarded time while $h_+(u)$ and $h_{\times}(u)$ are
plus polarization and cross polarization, respectively. For simple,
assuming the gravitational plane wave with a single frequency
$\omega$, its plus and cross polarizations are given by
$h_+(t-z)=A_+\cos(\omega(t-z)+\phi_+)$ and
$h_{\times}(t-z)=A_{\times}\cos(\omega(t-z)+\phi_{\times})$,
where $\phi_{+}$ and $\phi_{\times}$ are phases. Thus, the gravitational
plane wave is a ripple along the $z$ axis in spacetime propagating at
the speed of light. The plus and cross polarizations deform spacetime
in quadrupolar patterns. Since the Cartesian coordiantes are used to
describe the spacetime with gravitational plane wave, it may be influence
the PE spectrum band where the vibrational normal modes are linear
functions of Cartesian coordiantes. Furthermore, the metric tensor
of the gravitational plane wave deifned in Equation \eqref{eq:plane-grav-wave-metric}
is time-dependent. In this case, the whole EOMs become a set of coupled
nonlinear equations, which therefore require an iterative process to obtain
solution. This implies that the numerical treatment is highly complicated.
It can be expected that, we have to derive new coupled EOMs to overcome
this problem. 

Finally, it should be mentioned that the present framework in curved
spacetime differs from the so-called black hole chemistry \cite{kub17:063001}
completely in substance. Despite its nominal resemblance to the present
work, black hole chemistry is an thermodynamic framework for the properties
of black holes theirselves through the perspectives of chemical-thermodynamic
analogies. In black hole chemistry, the cosmological constant is treated
as a thermodynamic pressure, leading to a rich phase structure where
black holes exhibit behaviors analogous to conventional matter and its
thermodynamic properties \cite{kub17:063001}, such as van der Waals
fluids, phase transitions, triple points, and so forth. The mass of the
black hole is identified not as internal energy, but as chemical enthalpy,
while a newly defined thermodynamic volume allows for the formulation
of equations of state and the exploration of mechanical work. It has
revealed profound connections between gravity, thermodynamics, and
chemistry, opening new ways for understanding quantum gravity and
emergent spacetime. Therefore, black hole chemistry is essentially the
physics on black hole with chemical and phase thermodynamics as its
tool \cite{kub17:063001}. Since chemistry is the interplay of matter
and a vast subject with many diverse applications, black hole chemistry
may prove to be just as multi-faceted.

\subsection{Discussions\label{sec:mol-fiel}}

Based on present numerical results, now we must discuss on quantum
molecular dynamics in curved spacetime, as summarized by Table
\ref{tab:discuss-future-theory}. First of all, as given above in
Section \ref{sec:results-diss}, since the present theoretical framework
is not spacetime and Lorentz covariant, the next goal might be to
overcome this problem in falt space leading to chemistry of high-speed
molecules based on QFT \cite{zha25:20397} leading to chemical dynamics
with remarkably large velocity, such as chemistry on stars S62 and
S4714 with $\sim8\%$ the speed of light \cite{bou17:2151,lor20:186,lor22:209}.
It can be expected to possess asymptotic character, that is it is
applicable to the transition region in the vicinity of celestial bodies
where plasma gradually evolves into molecular systems. Next, we must
qualitatively examine the relevant scale of QFT in curved spacetime
and its matching degree with molecular processes. We would like mention
that at this case the molecular system with particles becomes quantum
field. The mass and
energy of the quantum field must be well below the Planck scale, namely
$m_{\mathrm{Planck}}\sim6\times10^{18}\;\mathrm{dalton}$ and
$E_{\mathrm{Planck}}\sim10^{28}\;\mathrm{eV}$, ensuring that the
back-reaction of the field on the spacetime geometry is small enough.
Compton wavelength of the particle in the system must be smaller than
the curvature radius making the concept of a localized particle possible.
Moreover, spatial scales of the system must be much larger than the
Planck length $l_{\mathrm{Planck}}\sim10^{-25}\;\mathrm{\AA}$, while
the spacetime changes slowly compared to the characteristic time scale
of the quantum field. Rapid changes in geometry can lead to excessive
particle creation violating the limitation for chemistry (physically,
called semi-classical approximation). Obviously, motions of the molecular
system (that is quantum field now) satisfy all of the above restrictions.

Second, it should be noted that the approximation, where geometry of
the spacetime does not change rapidly, effectively ensure the persistence
of the one-to-one mapping relationships illustrated in Figures
\ref{fig:reax-spacetime} and \ref{fig:map-slice-conf}. This is important
in studying chemical dynamics in curved spacetime where no particle
creation or annihilation. Furthermore,
molecular dynamics based on QFT reformulates the motions of different
particles into the dynamics of various types of quantum fields, while
the second quantization of these fields conversely allows a return to
the description of particle behavior. By the perspectives of field,
coupling between matter fields and gravitational field becomes possible
implying that one can consider the chemical dynamics beyond the concept
of absolute time. Next, we must turn to the separation of electrons and nuclei,
which is an important issue in molecular dynamics. The Born-Oppenheimer
approximation (BOA) and non-adiabatic methods beyond it are usually
adopted to separate chemical dynamics into electronic structure and
molecular dynamics, which are connected through the PES term. Similarly,
an adiabatic approximation and its non-adiabatic revisions also exist
in QFT due to the difference between fast and slow fields. By the
adiabatic approximation, a slowly changing field can be treated as a
fixed background. The other, fast-reacting quantum fields are then
studied as they move and fluctuate within this background. Similar
to the Higgs mechanism but unlike the BOA, the field associted with
electronic motions is considered as the background of the
field associated with nuclear motions. When the adiabatic
approximation breaks, the fields become strongly coupled with each
other, jumping energy levels, creating new particles, or dumping energy
back into the background field. 

Third, as the simplest application of QFT to molecular dynamics,
since the effects of nuclear spin are often igonred in considering
chemical dynamics, the quantum field associated with nuclei can
satisfy the generalized Klein-Gordon (KG) equation \cite{bir82:book,bar09:book,hac16:book}
\begin{equation}
\Big(\Box_g+m^2+\xi R\Big)\phi=V\phi,
\label{eq:kg-eq-qmd-0}
\end{equation}
where $R$ is the Ricci scalar curvature, $\xi$ a coupling constant,
and $\phi$ field function for nuclei motions. In Equation
\eqref{eq:kg-eq-qmd-0}, $\Box_g$ is the covariant d'Alembertian operator
satisfying \cite{bir82:book,bar09:book,hac16:book}
\begin{equation}
\Box_g\phi=\frac{1}{\sqrt{-g}}\partial_\mu
\Big(\sqrt{-g}g^{\mu\nu}\partial_\nu\phi\Big),
\label{eq:kg-eq-qmd-1}
\end{equation}
where $g$ is the determinant of the metric tensor $g_{\mu\nu}$.
Of course, in addition to problematic ``negative probability'' densities,
several theoretical problems arisen from the generalized KG equation
need more work to resolve. For instance, introduction of the term of
$V$ requires a more general covariant form of Equation \eqref{eq:kg-eq-qmd-0}.
To this end, the interaction $V$ should be introduced via minimal coupling
by replacing the ordinary derivative with the covariant derivative. In
this context, the use of $\partial_{\mu}\to\nabla_{\mu}=\partial_{\mu}-\mi A_{\mu}$
makes the interaction become potential field $A_{\mu}$. Moreover, the
generalized KG equation as given by Equation \eqref{eq:kg-eq-qmd-0} is
not renormalizable which limits its use to effective theories or external
fields.

Finally, it is interesting to give a perspective toward an experimental proposal.
To this end, we would like to mention that the dynamics and kinetics
of reactions on flat surfaces are theoretical foundations of catalytic
engineering. The idealization of flat surface makes numerical simulations
tractable and providing a clear framework for interpreting experiments.
However, it is increasingly revealing its limitations in the age of
nanotechnology, where nanostructured catalysts are characterized by
pronounced curvature, such as nanoparticles, nanotubes, and porous
frameworks. These curved surfaces create unique environments that
fundamentally alter chemical dynamics that are entirely absent in
flat-surface models. It is precisely this critical gap architectures
where our work becomes indispensable. For example, Figure \ref{fig:reax-spacetime}
can be considered as a model for the simplest reaction H + H$_2$ that
has been adsorbed on a curved surface. The present theory moves beyond
the constraint of planarity to account for the influence of surface
curvature on chemical dynamics and provides the essential tools to
predict catalytic behavior in real-world of nanoscale systems. This
advancement opens the door to the rational design of next-generation
catalysts with tailored geometries for unprecedented activity and
selectivity.

\section{Conclusions\label{sec:con}}

The present work proposes a primitive theory on chemical dynamics in
curved spacetime to introduce gravitational field in chemistry from
the absolute-sapce and universal-time viewpoint of Galileo.
In particular, the present goal is quantum molecular dynamics in
Schwarzschild spacetime which is a spherically symmetric solution to
Einstein field equations containing a central black hole at the origin.
One can propose a framework for describing quantum dynamics in Newtonian
gravity assuming a flat spacetime and introducing the gravitational
potential as a perturbative term to the PES. However, this method is
unable to study dynamics with slightly stronger gravitational field
where the spacetime is curved. In this work, noting the importance
role of the metric tensor in deriving the KEO, by revising the KEOs
we explore dynamics in curved spacetime for (1) the H + H$_2$ reaction,
(2) the H$_2$ + H$_2$ scattering, (3) dissociative chemsorption of
H$_2$O on Cu(111), (4) the spectrum band of anthracene cation, and
(5) the geometric phase in surface scattering. Extensive calculations
predict that the curved spacetime makes dynamical properties, such as
reaction or scattering probability and spectrum band, decrease abruptly
to zero as the gravitational strength increases; meanwhile, the geometric
phase is hardly influenced by the curvature of the spacetime. These numerical
results clearly indicate role of strong gravitational field in molecular
properties. Finally, the present theory represents only a preliminary
framework and thus incorporates several assumptions, which inevitably
introduce a considerable degree of approximation. To overcome these
limitations, we further propose potential strategies and disucss future
plans for addressing these assumptions in subsequent work.

\section*{Supplementary Material}

The Supporting Information files are available free of charge at
https://www.doi.org/XXXX. In the first Supporting Information file, we
give details of the present theory and numerical calculations on the
benchmarks. In the second Supporting Information file, we give details
of the vibrational normal coordinates of anthracene radical cation.

\section*{Interest Statement}

The authors declare that they have no known competing financial interests
or personal relationships that could have appeared to influence the work
reported in this paper.

\section*{Data Availability}

All data have been reported in this work. The data supporting this article
have been included as part of the Supplementary Information.

\section*{Acknowledgments}

The financial supports of National Natural Science Foundation of China
(Grant No. 22273074) and Fundamental Research Funds for the Central
Universities (Grant Nos. 2025JGZY34 and 2025KCW017) are gratefully acknowledged. The
authors also thank the {\it Centre National de la Recherche Scientifique}
(CNRS) International Research Network (IRN) ``MCTDH'' which is a
constructive platform for discussing quantum dynamics. The authors are
grateful to the authors of the PES of H$_2$O/Cu(111) in Dalian Institute
of Chemical Physics, Chinese Academy of Sciences for providing their
subroutines. The authors are also grateful to anonymous reviewers for
their thoughtful suggestions.

\clearpage
\begin{sidewaystable}
\caption{
Main assumaptions of the present theory which presents the molecular
motions in curved spacetime. The second column gives target of each
assumption. The third column describes the assumptions. The rightmost
column gives remarks of these assumptions. The present six points are
actually arisen from three assumptions. First, the metric tensor is a
typical solution of Einstein equation which is further transferred and
implemented by the ADM equations while the 4D spacetime is $[3+1]$
decomposed. Second, the gravity only provides a geometric background
of spacetime. Third, the geometric characteristics of the 3D slice
$\Sigma_t$ remain time-invariant throughout the reaction process. This
is arisen from the assumption that curved spacetime only affects the
geometry.
}%
\begin{tabular}{lclclcr}
\hline
No. &~~& Target &~~& Description &~~& Remark \\
\hline
1 && metric && classical (non-quantized)  && A typical solution of Einstein  \\ 
  &&        && Lorentzian metric $g_{\mu\nu}$ && equation is required  \\
2 && gravity && not a dynamical field 
&& It provides a geometric background of spacetime \\
3 \footnote{The third and fourth assumptions encode that curved
spacetime only affects the geometry but not the molecular motions.
\label{foot:assu3-4}}
&& coordinates transfer && local Lorentz invariance
\footnote{Local Lorentz symmetry holds at every point in spacetime.}
&& Theories in curved space reduce to  \\
 &&  &&  && familiar flat-space theories at each point \\
4 \textsuperscript{\ref{foot:assu3-4}} && generalization && minimal coupling 
&& Derivatives are replaced with covariant \\
&& && && derivatives while metric is inserted
\footnote{The flat-space equations are generalized to those in curved
space by replacing derivatives with covariant derivatives and inserting
the metric.}  \\
5 \footnote{This assumption indicates that particle becomes observer-dependent
rather than observer-independent as usually expected.}
&& time && universal time && It is acceptable due to the Eulerian observer \\
6 \footnote{This assumption is arisen from the second assumption for
gravity.}
&& counteraction && no influence of molecule on the geometry
&& Semiclassical Einstein equation for the \\
&& && && back reaction of molecular  \\
&& && && motions is not considered \\
7 \footnote{This assumption is employed only for the present numerical demonstration.}
&& spacetime 
&& Schwarzschild
&& It is a spherically symmetric, uncharged, \\
&& && && and non-rotating static spacetime. \\
\hline
\end{tabular}
\label{tab:assum-curved}
\end{sidewaystable}

\clearpage
\begin{table}
\caption{
Values of $\varrho=r_s/r$ in the revised KEOs for the ML-MCTDH calculations
to show the present theory, together with the present benchmarks. These
benchmarks include (1) the H + H$_2$ reaction, (2) the H$_2$ + H$_2$
scattering, (3) dissociative chemsorption of H$_2$O on Cu(111), (4) the
spectrum band of anthracene cation, and (5) the Berry phase in the nuclear
wave function. These systems represent typical processes in chemical
dynamics. The second column gives various values of $\varrho$ used in
this work. The third column presents the systems using the corresponding
$\varrho$ values. The rightmost column presents remarks for values of
$\varrho$. We refer the reader to Reference
\cite{sch12:book,mis17:book,car19:book} for details.}
\begin{tabular}{lclcccccccr}
\hline
No. &~& Value of $\varrho$ &~& \multicolumn{5}{c}{Systems
\footnote{Symbol $\checkmark$ means that the value of $\varrho$ is
used for the respective system. Symbol $\times$ represents the opposite
scenario.}} &~& Remark \\ \cline{5-9}
    &~&  &~& H + H$_2$ \footnote{The H + H$_2$ and H$_2$O/Cu(111) systems
are typical reactions in the gas phase and on the surface, respectively.
\label{foot:reax-def}}
& H$_2$ + H$_2$ \footnote{The H$_2$ + H$_2$ and anthracene systems
are typical bound systems but offering rich chemical insights.\label{foot:non-reax-def}}
& H$_2$O/Cu(111) \textsuperscript{\ref{foot:reax-def}}
& anthracene \textsuperscript{\ref{foot:non-reax-def}} 
& Berry phase \footnote{Section S-IV of the SI file presents many more interpretations and discussions.}
&~&  \\
\hline
1  && $0.60$ && $\checkmark$ & $\times$ & $\times$ & $\checkmark$ & $\times$ && strong gravity field
\footnote{At strong gravity field with $\varrho\to1$, one cannot observe
an object precisely, as given in the first SI file.}  \\
2  && $0.50$ && $\checkmark$ & $\checkmark$ & $\times$ & $\checkmark$ & $\checkmark$ &&
noticeable gravity field  \\
3  && $0.40$ && $\checkmark$ & $\times$     & $\checkmark$ & $\times$ & $\times$ &&
noticeable gravity field  \\
4  && $0.30$ && $\checkmark$ & $\checkmark$ & $\checkmark$ & $\checkmark$ & $\checkmark$ &&
noticeable gravity field  \\
5  && $0.20$ && $\checkmark$ & $\times$     & $\times$ & $\times$ & $\times$ &&
moderate gravity field    \\
6  && $0.10$ && $\checkmark$ & $\checkmark$ & $\checkmark$ & $\checkmark$ & $\checkmark$ &&
around a stellar-mass black hole, \\
   &&        && & & & & && say Cygnus X-1
\footnote{Cygnus X-1 is one of the brightest X-ray sources and a firmly
established stellar-mass black hole.}  \\
7  && $0.05$ && $\checkmark$ & $\checkmark$ & $\checkmark$ & $\checkmark$ & $\checkmark$ &&
moderate gravity with \\
   &&        && & & & & && noticeable relativistic effects \\
8  && $0.03$ && $\checkmark$ & $\checkmark$ & $\checkmark$ & $\times$ & $\checkmark$ &&
around massive object \\
   &&        && & & & & &&  with weak gravity \\
9  && $0.01$ && $\checkmark$ & $\checkmark$ & $\checkmark$ & $\checkmark$ & $\checkmark$ &&
around massive object, \\
   &&        && & & & & && say neutron star
\footnote{A neutron star is an extremely dense stellar remnant with strong magnetic fields.} \\
10 && $0.00$ && $\checkmark$ & $\checkmark$ & $\checkmark$ & $\checkmark$ & $\checkmark$ &&
flat space without gravity, \\
   &&        && & & & & && say around the earth \\
\hline
\end{tabular}
\label{tab:num-varrho-values}
\end{table}

\clearpage
\begin{table}
 \caption{%
Parameters in the present 96D + 2D Hamiltonian model for the molecule-surface
system. Parameters of $\hat{H}_{\mathrm{surf}}'$ given in Equations
\eqref{eq:model-ham-002} and \eqref{eq:model-ham-003} were previously
optimzed for the Cu(100) surface \cite{men21:2702}. The second column
gives the parameters in defining the models. The third column gives
values of these parameters. The frequency parameters $\eta$ and $\xi$
of the CHO model represent conditions of the processes, such as the
nozzle and target temperatures in the surface scattering. The
fourth column gives expression in which the parameter appear. The
rightmost column gives remarks on these parameters.
}%
 \begin{tabular}{lclclclcr}
  \hline
No. &~& Parameter  &~& Value &~& Expression &~& Remark  \\
\hline
1 && $m_1$ && $16.000$ dalton && Equation \eqref{eq:3d-cho-hamiltonian-00} 
&& mass of the atom in molecule \\
2 && $m_2$ && $12.000$ dalton && Equation \eqref{eq:3d-cho-hamiltonian-00}
&& mass of the atom in molecule \\
3 && $m_3$ && $63.546$ dalton && Equation \eqref{eq:3d-cho-hamiltonian-00} 
&&  mass of surface atom \\
4 && $\omega_2$ && $500.0\;\mathrm{cm}^{-1}$
\footnote{They are set to be typical values for a poly-atomic molecule
adsorbed on a metal surface.\label{foot:vib-frquen}}
&& Equation \eqref{eq:3d-cho-hamiltonian-00} && force constant of the adsorption \\
5 && $\eta$ && $\omega_1^2/\omega_2^2$ && Equation \eqref{eq:3d-cho-hamiltonian-00}
&& function of nozzle temperature \\
6 && $\xi$ && $\omega_3^2/\omega_2^2$ && Equation \eqref{eq:3d-cho-hamiltonian-00}
&& function of surface temperature \\
7 && $k_b$    && $1000.0$ cm$^{-1}$ \textsuperscript{\ref{foot:vib-frquen}}
&& Equation \eqref{eq:model-ham-001} && force constant of the bond \\
8 && $k_R$    && $1.000\times10^{-2}$ hartree 
\footnote{The value of $k_R$ is set to be a perturbation.}
&& Equation \eqref{eq:model-ham-001} && force constant along the $R$ coordinate \\
9 && $b_0$    &&  $2.267$ bohr \textsuperscript{\ref{foot:vib-frquen}}
&& Equation \eqref{eq:model-ham-001} && bond length \\
10&& $R_0$    &&  $0.397$ bohr
\footnote{The value of $R_0$ is set to be largely smaller than the
distance $d$ between surface atoms.}
&& Equation \eqref{eq:model-ham-001} && radius of the closed loop \\
11&& $\{Q_{i\mu}^{(0)}\}_{i=0}^{29}$, $\mu=x,y,z$ && ---
\footnote{Values of parameters in $\hat{H}_{\mathrm{surf}}'$ were optimized
for the Cu(100) surface \cite{men21:2702}.\label{foot:opt-para-co-cu}}
&& Equation \eqref{eq:model-ham-002} && optimized coordinates of the $i$-th atom \\
12 && $d$ && $4.830$ bohr
\footnote{The value of $d$ was computed by geometry optimization for
the Cu(100) surface \cite{men21:2702}.}
&& Equation \eqref{eq:model-ham-002} && lattice constant \\
13 && $k_Z$ && $1.100\times10^{-2}$ hartree \textsuperscript{\ref{foot:opt-para-co-cu}}
&& Equation \eqref{eq:model-ham-002} && coupling strength the non-top atoms \\
14&& $D$ && $0.300$ hartree \textsuperscript{\ref{foot:opt-para-co-cu}}
&& Equation \eqref{eq:model-ham-003} && deep of the Morse potential \\
15&& $a$ && $0.390$ {\AA}$^{-1}$ \textsuperscript{\ref{foot:opt-para-co-cu}}
&& Equation \eqref{eq:model-ham-003} && width parameter of Morse well \\
\hline
\end{tabular}
\label{tab:98d-model}
\end{table}

\clearpage
\begin{sidewaystable}
\caption{
Problems of the present theoretical framework and their potential
solutions disucced in Sections \ref{sec:results-diss} and \ref{sec:mol-fiel}.
The second column gives the problems arisen from the present calculations
and assumptions given in Table \ref{tab:assum-curved}. The third column
simply describes source of the problem. The fourth column simply gives
respective solution. The rightmost column gives the section that provides
an explanations of this problem and solution. Since many of these problems
are open questions, respective solution is just a proposal. Here, let
$\{\mi t,r,\theta,\phi\}$
be the 4D spherical polar cooridnates of the Schwarzschild sapcetime
and let $\{\mi t,x,y,z\}$ be the 4D Cartesian coordinates set of the
sapcetime with plane gravitational wave.}
\begin{tabular}{lclclrcr}
\hline
No. &~~& Problem &~~& Source & Solution &~~& Section \\
\hline
1 && How about gravitational time  && gravitational field and & the present numerical reuslts
&& \ref{sec:relativity} and \ref{sec:results-diss} \\
  && dilation? && curved spacetime & on molecular processes && \\
2 && How about internal rotations? && assume $r$ to be large enough
\footnote{The system is far from center of the Schwarzschild spacetime making
atoms experience a uniform curvature.}
& the angular momenta of the KEO  && \ref{sec:results-diss} \\
3 && How about orbital rotations? && assume $\theta$ and $\phi$ to be fixed
\footnote{The system does not undergo orbital rotation ensuring its low velocity.}
& angular elements of the metric for && \ref{sec:results-diss} \\
&& && & orbital angular momenta of the system && \\
4 \footnote{These two problems might be considered together.\label{foot:36-same}}
&& How about coupling between  && separate them and use the
& new EOMs with such coupling
\footnote{For example, the new EOMs contain Schr{\"o}dinger-type
equation, ADM equations, and their coupling. See also the sixth
problem in this table.} 
&& \ref{sec:mol-fiel} \\
  && the system and spacetime?   && concept of absolute time &  && \\
5 && How about absolute time?    && use Schr{\"o}dinger-type EOM
\footnote{It is possible due to the $[3+1]$ decomposition, however,
which does not require the concept of absolute time.}
& QFT with covariant EOMs  && \ref{sec:mol-fiel} \\
  &&  &&  & under Lorentz transformation && \\
6 && How about other spacetime? && put system in a spherically & KEO based on other metric
&& \ref{sec:results-diss} \\
  && &&  symmetric spacetime
\footnote{It is Schwarzschild spacetime and the simplest solution of the ADM equations.} 
& or QFT in curved space \footnote{See, for example, References
\cite{bir82:book,bar09:book,hac16:book}.} && \\
7 \textsuperscript{\ref{foot:36-same}}
&& How to choose metric? && separate system and sapcetime
\footnote{See also the third problem in this table.} &
igonore coupling between && \ref{sec:mol-fiel}   \\
  &&  &&  & system and sapcetime
\footnote{Separation of system and spacetime may lead to approximation
due to considerable divergence in their dynamics.} &&  \\
\hline
\end{tabular}
\label{tab:discuss-perspe}
\end{sidewaystable}

\clearpage
\begin{sidewaystable}
\caption{
Same as Table \ref{tab:discuss-perspe}, except for the theoretical
problems of chemistry in curved space and their potential solutions
or answers disucced in Sections \ref{sec:results-diss} and
\ref{sec:mol-fiel}. We refer the reader to Figure S-2 of the
first SI file for further discussions on these problem.}
\begin{tabular}{lclclrcr}
\hline
No. &~~& Problem &~~& Source & Solution &~~& Section \\
\hline
1 && Can QFT in flat space   && chemistry of fast   & covariant EOMs for molecules
\footnote{The molecular EOMs must be covariant under Lorentz transformation.}
  && \ref{sec:mol-fiel} \\
  && be applied to chemistry? && moving molecules
\footnote{It means chemistry of molecules with remarkably large velocity,
such as chemistry on stars S62 and S4714 with $\sim8\%$ the speed of light
\cite{bou17:2151,lor20:186,lor22:209}.}
  &   &&   \\
2 && Can QFT in curved space  && chemistry based on QFT & 
$m_{\mathrm{mol}}\ll m_{\mathrm{Planck}}\sim10^{18}$ dalton, 
&& \ref{sec:mol-fiel} \\
&& be applied to chemistry? &&  &
$E_{\mathrm{mol}}\ll E_{\mathrm{Planck}}\sim10^{28}$ eV, && \\
&& && & $l_{\mathrm{mol}}\gg l_{\mathrm{Planck}}\sim10^{-25}$ \AA &&  \\
3 && Can electrons and nuclei && convenience of the & adiabatic approximations 
\footnote{Typical examples are (1) background field method of QFT in curved
spacetime, (2) effective field theory, and (3) non-relativistic limit.}
  && \ref{sec:results-diss} \\
  && be separated in QFT? && Born-Oppenheimer &  &&  \\
  && && approximation
\footnote{The Born-Oppenheimer approximation allows for separation of
electronic structure and quantum molecular dynamics simplifying the
overall calculations.}  &  &&  \\
4 && Can a unified theory exist  && distinction of electromagnetic & unified theory of fields
\footnote{It would seamlessly blend the world of quantum particles with
the cosmos. However, it has not been found and become a huge challenge
in physics.}
&& \ref{sec:mol-fiel} \\
  && for molecule and spacetime? && and gravitational fields  & && \\
\hline
\end{tabular}
\label{tab:discuss-future-theory}
\end{sidewaystable}

\clearpage
 \section*{Figure Captions}

\figcaption{fig:reax-spacetime}{%
Decomposition of the 4D spacetime with metric tensor
$g_{\mu\nu}$, which can be foliated into a family of non-intersecting
spacelike 3D surfaces $\Sigma_t$, called time slices. Subfigures (a)
and (b) illustrate the decompositions of flat and curved spacetime,
respectively. The vertical axis represents the time axis. For easily
understanding, a reaction process of H + H$_2$ is shown in each slice
as a typical example. These 3D surfaces arise, at least locally, as
the level surfaces of a scalar function $t$ which can be interpreted
as a global time function. From $t$, one can define a 1-form as
$\Omega_{\mu}=\nabla_{\mu}t$ and then define the lapse function
$\alpha$. By the unit normal vector $n^\mu$ to the slice $\Sigma_t$,
one can introduce the spatial metric tensor $\gamma_{\mu\nu}$. To
introduce $t^{\mu}$ that connects the spatial coordinates of adjacent
slices, the spatial shift vector $\beta^{\mu}$ is also defined to represent
the offset of spatial coordinates relative to the normal vector. With
$n^{\mu}$ and $\gamma^{\mu\nu}$, the extrinsic curvature $K_{\mu\nu}$
is defined by $K_{\mu\nu}=-\gamma_{\mu}^{\rho}\gamma_{\nu}^{\sigma}\nabla_{\rho}n_{\sigma}$.
Having the time basis vector, the geometric characteristics of each
slice are given by $(\gamma_{ij},K_{ij})$.
}%

\begin{figure}[h!]
 \centering
  \subfigure[\quad Reaction in curved spacetime]{%
   \includegraphics[height=22cm]{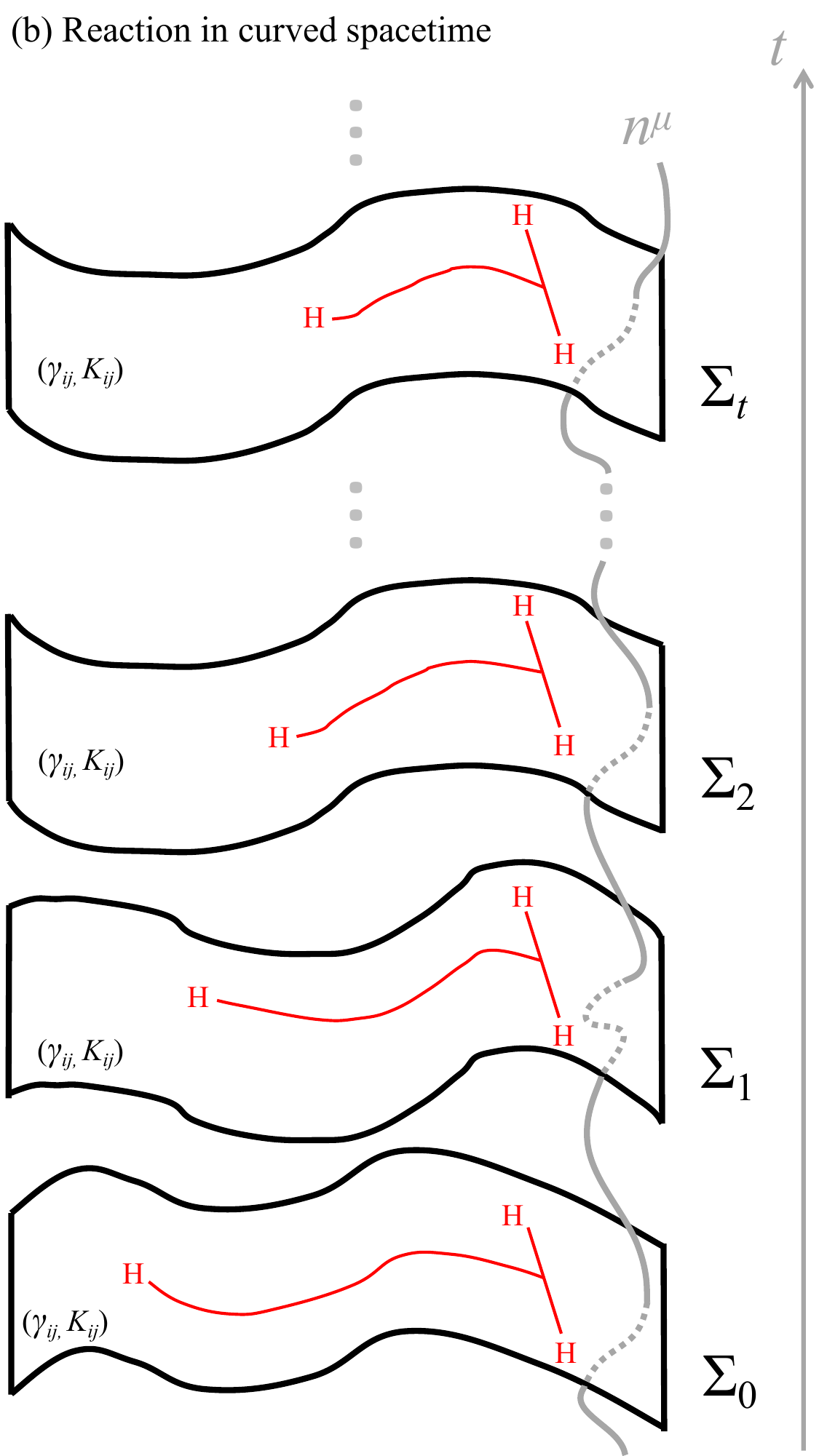}}    
    \caption{\figfoot}
     \label{fig:reax-spacetime}
      \end{figure}

\figcaption{fig:map-slice-conf}{%
Mapping relation between 3D slice and $3N$-dimension configuration
space, where subfigures (a) and (b) give the cases of flat and curved
slices, respectively. Similar to Figure \ref{fig:reax-spacetime}, the
reaction process of H + H$_2$ is shown as a typical example. The left
panel represents either flat or curved slice, while the right panel
represents $3N$-dimensional configuration space. In the configuration
space, the coordinates of $N$-atomic molecular system as given in the
left panel by red words are represented by one point, as shown by the
blue point in the right panel. The gray dashed lines connects atoms (in
red) and the representation point (in blue).
}%

\figcaption{fig:ml-tree-98d}{
The ML-MCTDH wave function structure (called ML-tree structure) for the
present quantum dynamics calculations of (a) the anthracene radical
cation, (b) the H$_2$O/Cu(111) system, (c) main part of the 96D + 2D
surface scattering model, and (d) bath part of the 96D + 2D surface
scattering model. The number of SPFs
are also given. The numbers of primitive basis sets to represent SPFs
of the deepest layer are given next to the lines connecting with the
squares. In subfigure (a), the maxima depth of the ML-trees
are six layers, and the first layer separates the $66$ vibrational
normal coordinates from the discrete electronic DOF. The vibrational
modes with a star are the ones which are included
in the reduced dimensional Hamitonian that was used for fast ML-MCTDH
calculations \cite{men13:014313}. Reprinted with permission from
Reference \cite{men13:014313}. Copyright 2013 American Institute of
Physics. In subfigure (b), the maxima depth is five layers and the
first layer separates the three coordinates of the H$_2$O COM from
the coordinates of the inner motions of H$_2$O. Reprinted with
permission from Reference \cite{son22:6047}. Copyright 2022 American
Chemical Society. For the 96D + 2D ML-MCTDH calculations, the maximum
depth of the ML-tree is six layers, and the first layer separates the
moelcular coordinates from those of the surface atoms. For simple, subfigure
(c) only shows the first $35$ coordinates, while subfigure (d) shows
the other $63$ coordinates that should locate at the ML-tree structure
in the red box of subfigure (c).
}

\figcaption{fig:num-results}{%
Dynamics results of the present ML-MCTDH calculations for (a) the
H + H$_2$ $\to$ H + H$_2$ reaction probability, (b) the H$_2$ + H$_2$
scattering probability, (c) dissociative chemsorption propabilities of
H$_2$O on Cu(111), and (d) spectrum bands of the anthracene cation in
$\tilde{B}\;{}^2A_u$, where the KEOs are revised according to the metric
of the curved spacetime. These subfigures especially illustrate the
dependence on the parameter $\varrho=r_s/r$ which is defined in the
Schwarzschild metric, as given by Equation \eqref{eq:schd-metric-000-revi}.
The cyan, maroon, light yellow, light green, and violet lines represent
quantity curves for $\varrho$ values of $0.60$, $0.50$, $0.40$, $0.30$,
and $0.20$, respectively. In addition, the yellow, green, blue, and red
lines represent those for $\varrho$ of $0.10$, $0.05$, $0.03$, and $0.01$,
respectively. The black line is the quantity curve computed in flat spacetime
with $\varrho=0$. Subfigure (a) illustrates reaction probability of the
H + H$_2$ system as function of kinetics energy (in eV) of the impacting
H atom. For $\varrho>0.6$, the computed reaction probabilities are very
close to zero and hence ignored. Subfigure (b) illustrates scattering
probability of the H$_2$ + H$_2$ system as function of impacting energy
(in eV). As is well known, such scattering propability in flat spacetime
is equal to unit since no reaction occurs between two H$_2$ molecules.
Subfigure (c) illustrates dissociative chemsorption propability of H$_2$O
on Cu(111) as function of impacting energy (in eV) of water. Subfigure
(d) illustrates the spectrum bands of the anthracene cation in the
$\tilde{B}\;{}^2A_u$ state.
}%

\begin{figure}[h!]
 \centering
  \subfigure[\quad Spectrum of the anthracene cation in curved spacetime]{
   \includegraphics[width=18cm]{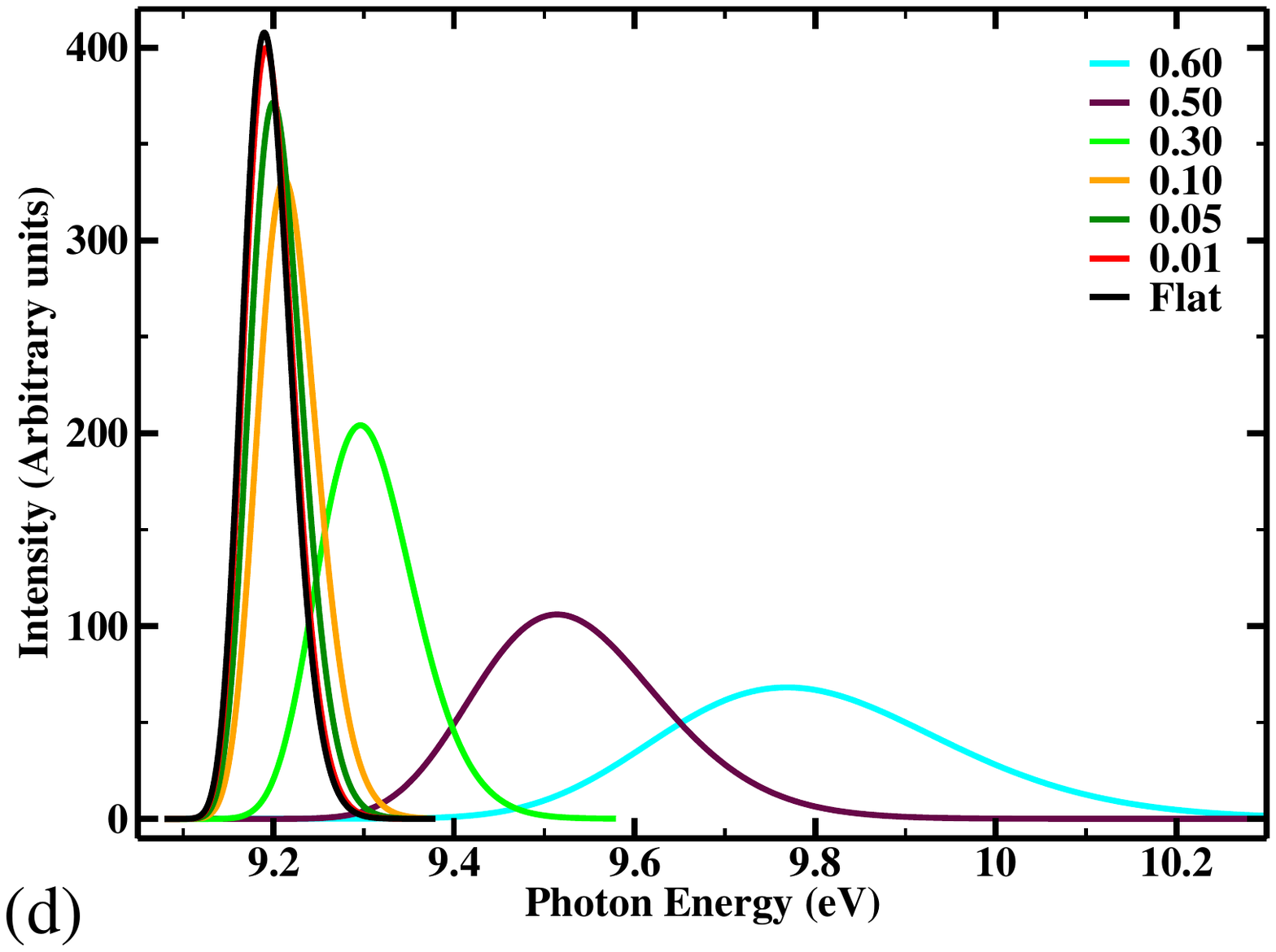}}
    \caption{\figfoot}
     \label{fig:num-results}
      \end{figure}

\figcaption{fig:berry-phase-var}{%
Same as Figure \ref{fig:num-results}, except for the Berry phase of the 98D
model for surface scattering. Here, the time interval over which the
system acquires a geometric phase of $\pi$ is shown. The light green,
yellow, green, blue, and red symbols represent those for $\varrho$ of
$0.30$, $0.10$, $0.05$, $0.03$, and $0.01$, respectively, while the
black symbols present the phase angles in flat space with $\varrho=0$.
The same curves of full time interval are given in the SI file.
}%

\figcaption{fig:norm-wavefunc}{
Comparsions of the square of the wave function norm $\Vert\Psi(t)\Vert^2
=\langle\Psi(t)\vert\Psi(t)\rangle$
at the cases with different values of $\varrho$ for (a) the H + H$_2$
$\to$ H + H$_2$ system, (b) the H$_2$ + H$_2$ system, and (c) the
H$_2$O/Cu(111) system, together with the autocorrection $C(t)=\langle\Psi(0)
\vert\Psi(t)\rangle$ of (d) the anthracene cation. The colored lines represent the
results computed by various $\varrho$ values. The cyan, maroon, light
yellow, light green, and violet lines represent quantity curves for
$\varrho$ values of $0.60$, $0.50$, $0.40$, $0.30$, and $0.20$,
respectively. In addition, the yellow, green, blue, and red lines
represent those for $\varrho$ of $0.10$, $0.05$, $0.03$, and $0.01$,
respectively. The black line is the quantity curve computed in flat
spacetime with $\varrho=0$. We must emphasize that, since a CAP is
employed in the propagations for these systems, the wave function
only satisfies conserved normalization in the early stage of
propagation. Once the CAP becomes active, conserved normalization
is lost. For this reason, we truncate the results at some time.
}

\begin{figure}[h!]
 \centering
  \subfigure[\quad Autocorrection of wave function of anthracene in curved spacetime]{
   \includegraphics[width=18cm]{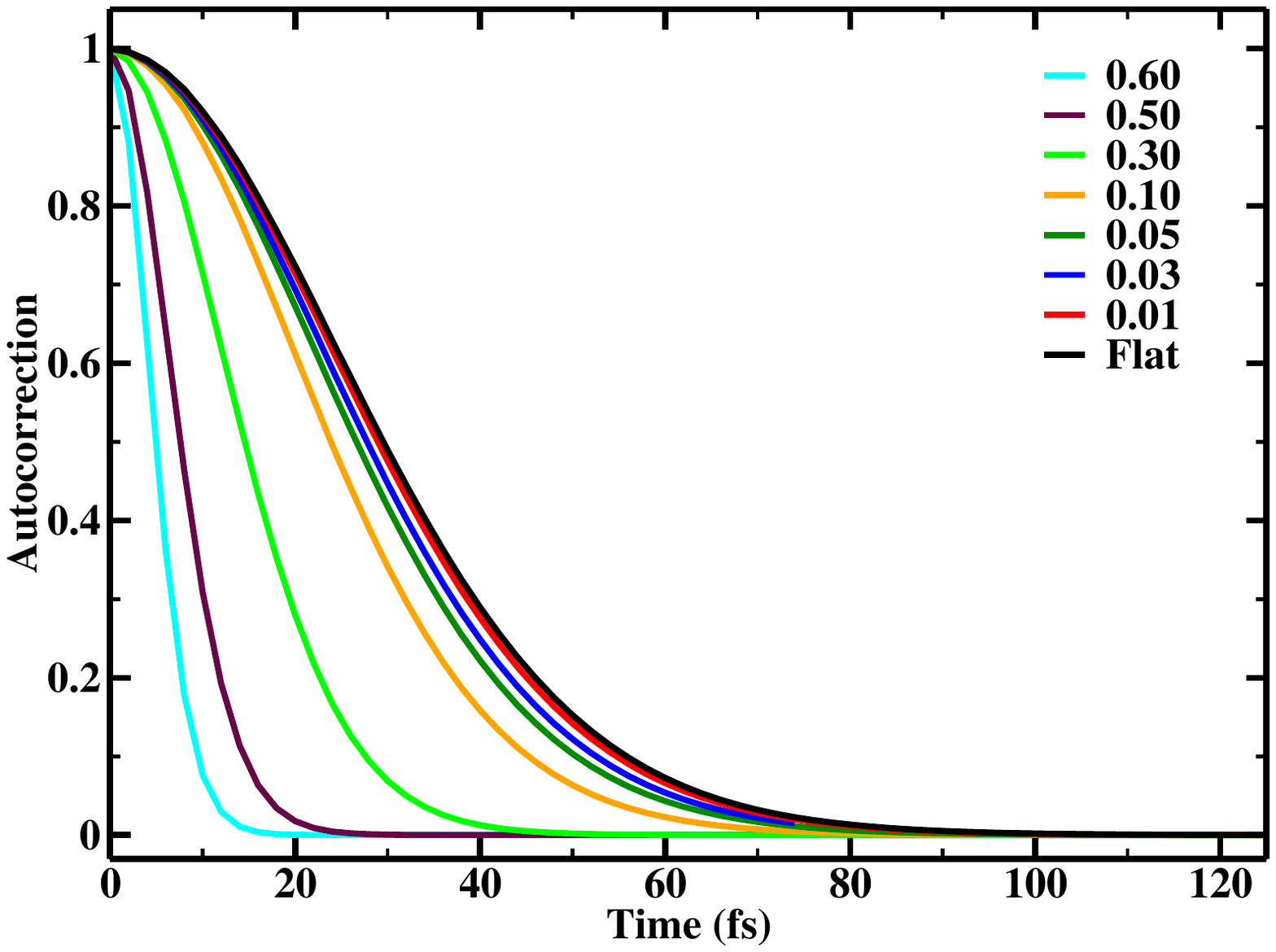}}
    \caption{\figfoot}
     \label{fig:norm-wavefunc}
      \end{figure}

\figcaption{fig:exp-proper}{
Comparsions of the time-dependent expectation values of the total energy
(in hartree) of (a) the H + H$_2$ $\to$ H + H$_2$ system, (b) the H$_2$
+ H$_2$ system, (c) the H$_2$O/Cu(111) system, and (d) the anthracene
cation, at different values of $\varrho$. These expectations are computed
by propagated wave function $\Psi(t)$ thorugh the expression $\langle
E\rangle(t)=\langle\Psi(t)\vert\hat{H}\vert\Psi(t)\rangle/\langle\Psi(t)\vert\Psi(t)\rangle$,
where $\hat{H}$ is the Hamiltonian operator at different cases. The
colored lines represent the results computed by various $\varrho$
values. The cyan, maroon, light yellow, light green, and violet lines
represent quantity curves for $\varrho$ values of $0.60$, $0.50$,
$0.40$, $0.30$, and $0.20$, respectively. In addition, the yellow,
green, blue, and red lines represent those for $\varrho$ of $0.10$,
$0.05$, $0.03$, and $0.01$, respectively. The black line is the
quantity curve computed in flat spacetime with $\varrho=0$. Here,
we must emphasize that, since a CAP is employed in the propagations
for these systems, the wave function only satisfies conserved normalization
$\langle\Psi(t)\vert\Psi(t)\rangle=\mathrm{constant}$ in the early
stage of propagation. Once the CAP becomes active, as illustrated by
Figure \ref{fig:norm-wavefunc}, conserved normalization is lost, and
$\langle E\rangle$ no longer satisfy conservation. For this reason,
we truncate the energy expectations at some time.
}

\clearpage
\begin{figure}[h!]
 \centering
  \subfigure[\quad Reaction in flat spacetime]{%
   \includegraphics[height=22cm]{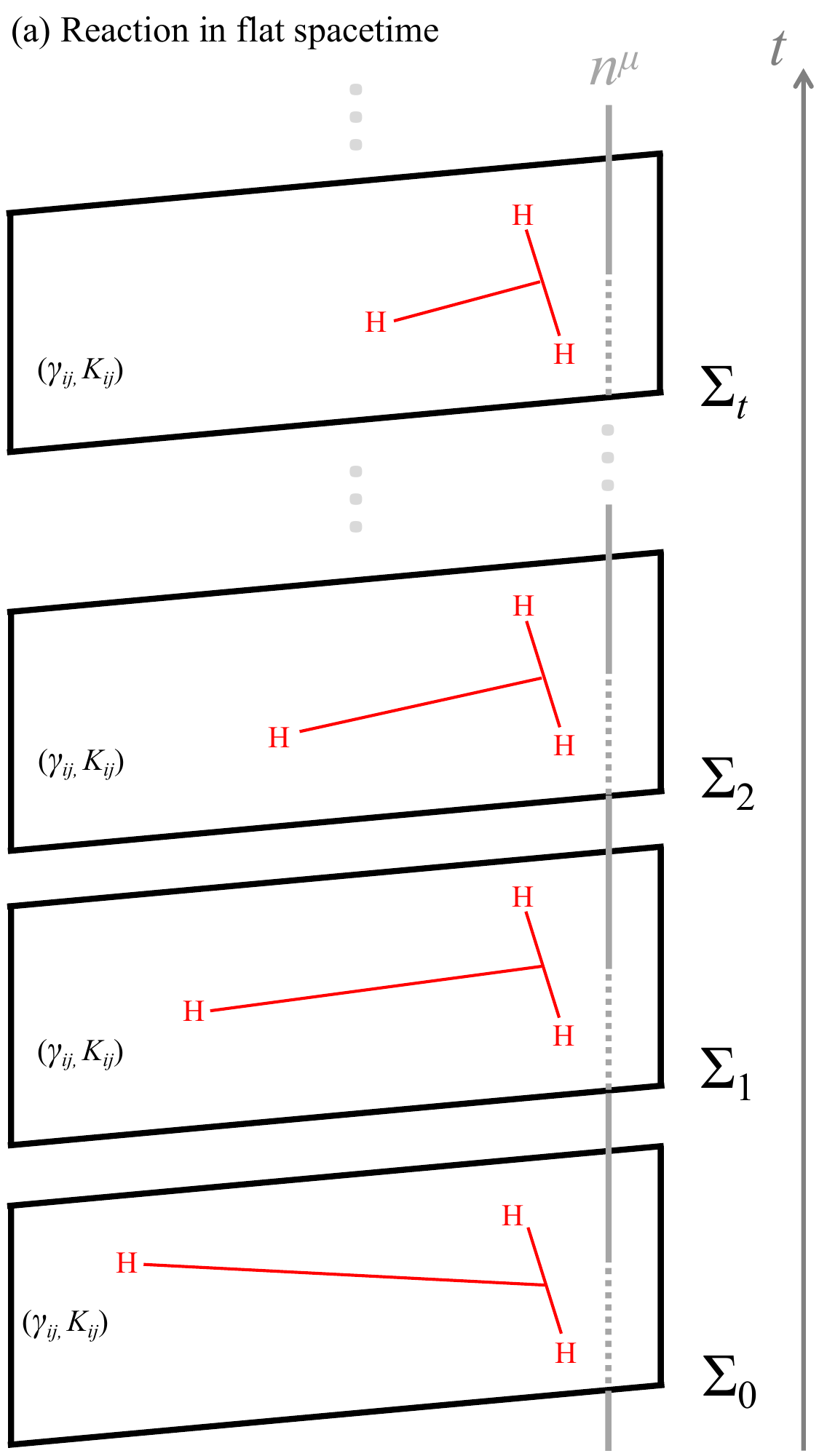}}
    \end{figure}
\clearpage

\clearpage
\begin{figure}[h!]
 \subfigure[\quad Mapping from flat spacetime to configuration space]{%
  \includegraphics[width=18cm]{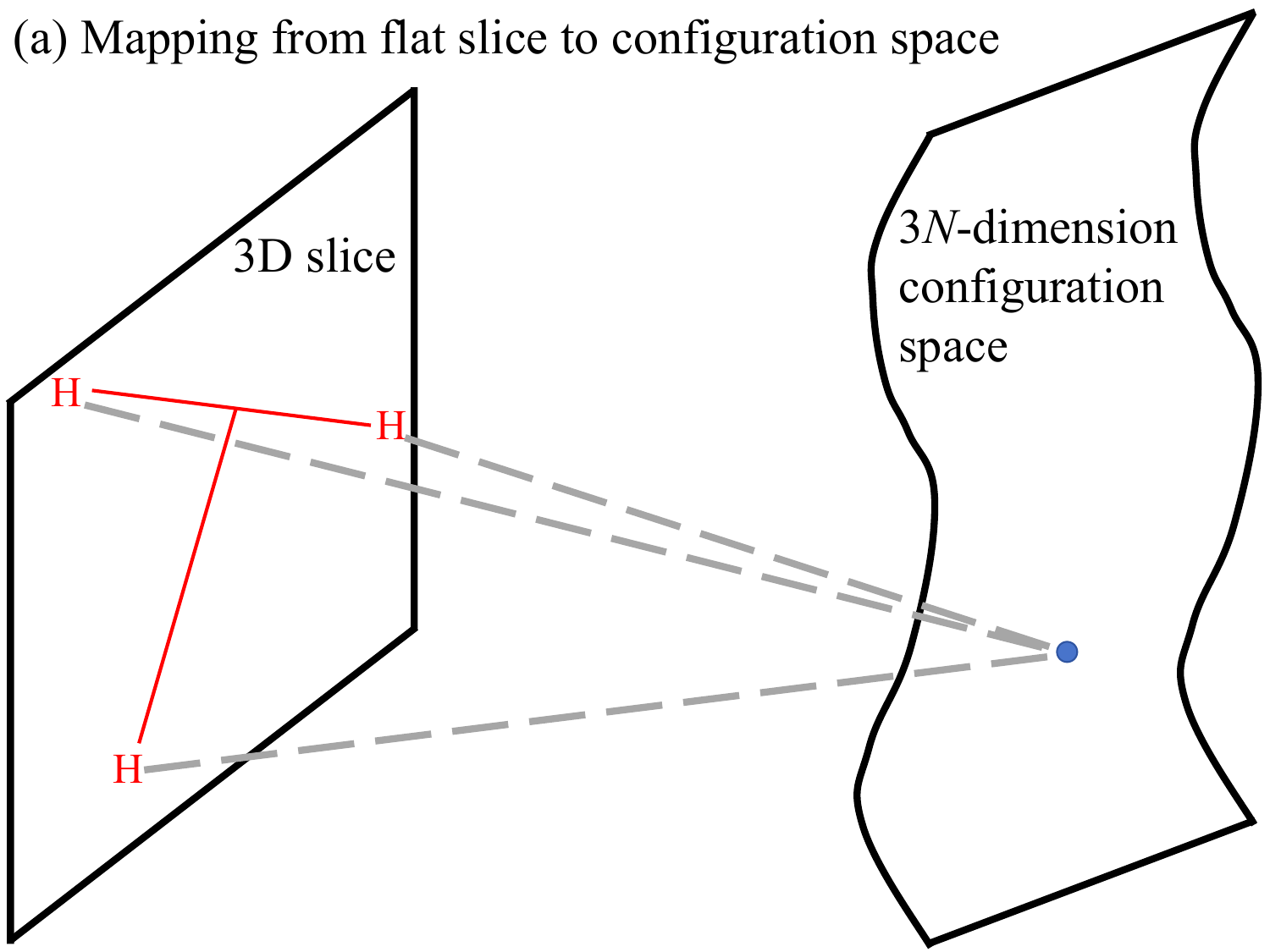}}
   \end{figure}
\clearpage

\clearpage
 \begin{figure}[h!]
  \centering
   \subfigure[\quad ML-tree of the 66D wave function for the anthracene cation]{
    \includegraphics[angle=90,height=24cm]{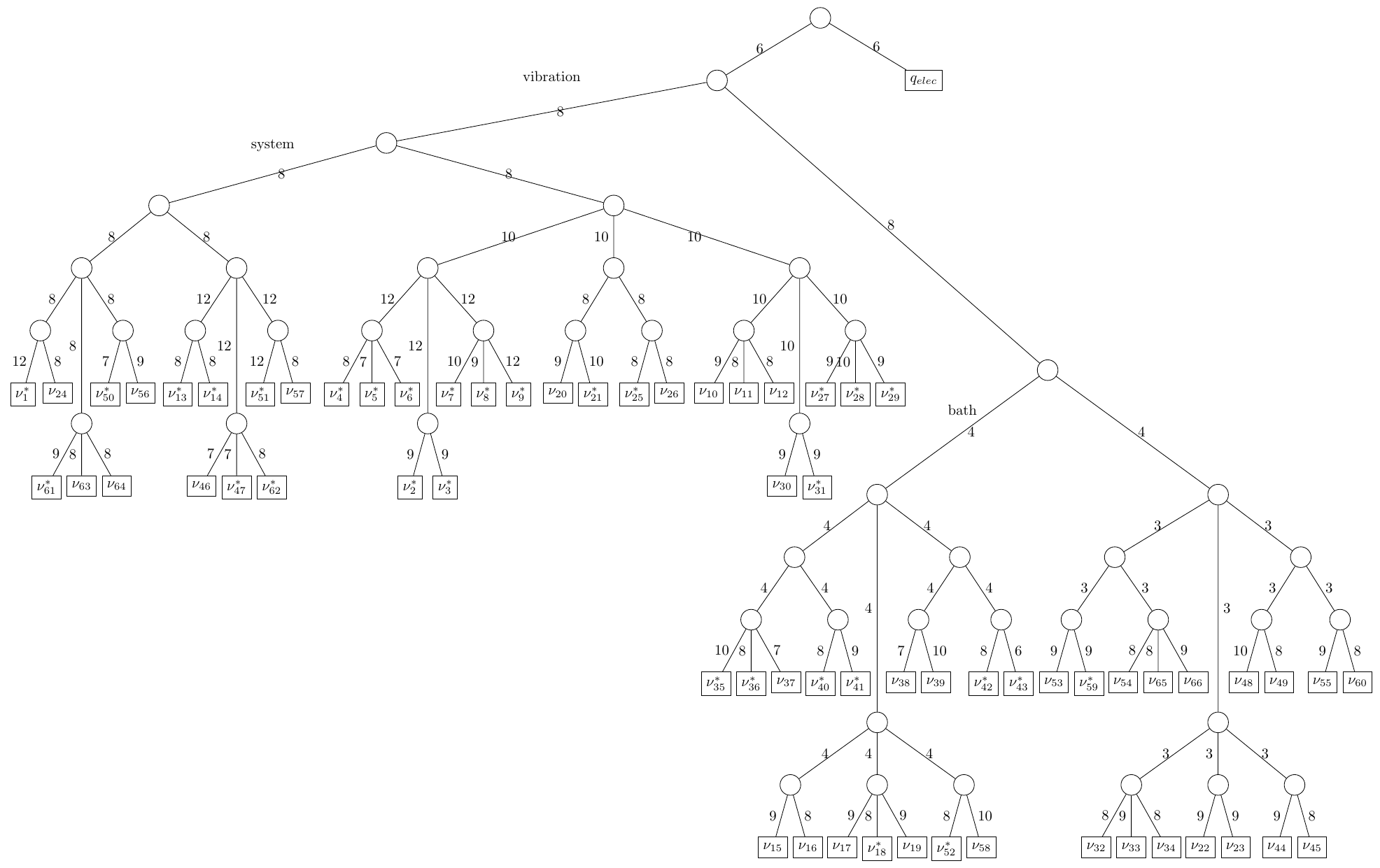}}
     \end{figure}
\clearpage     
 \begin{figure}[h!]
  \centering
   \subfigure[\quad ML-tree of the 9D wave function for the H$_2$O/Cu(111) system]{ 
    \includegraphics[width=18cm]{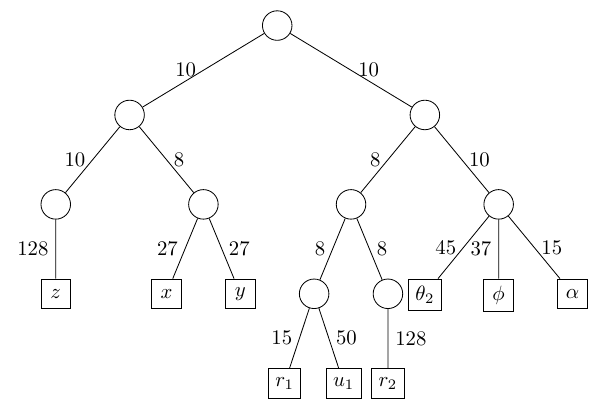}}
     \end{figure}
\clearpage
 \begin{figure}[h!]
  \centering
   \subfigure[\quad Main part of ML-tree of the 96D + 2D wave function]{%
    \includegraphics[width=23cm,angle=90]{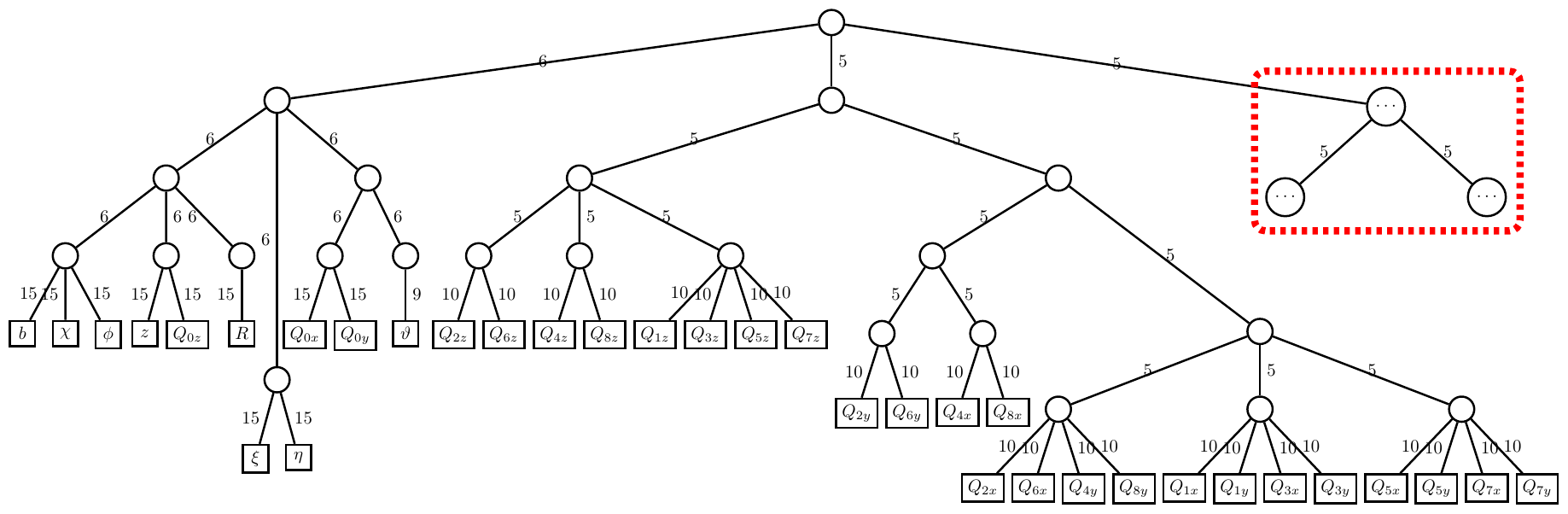}}
     \end{figure}
 \clearpage
       
\clearpage
\begin{figure}[h!]
 \centering
  \subfigure[\quad Reaction probability of H + H$_2$ in curved spacetime]{
   \includegraphics[width=18cm]{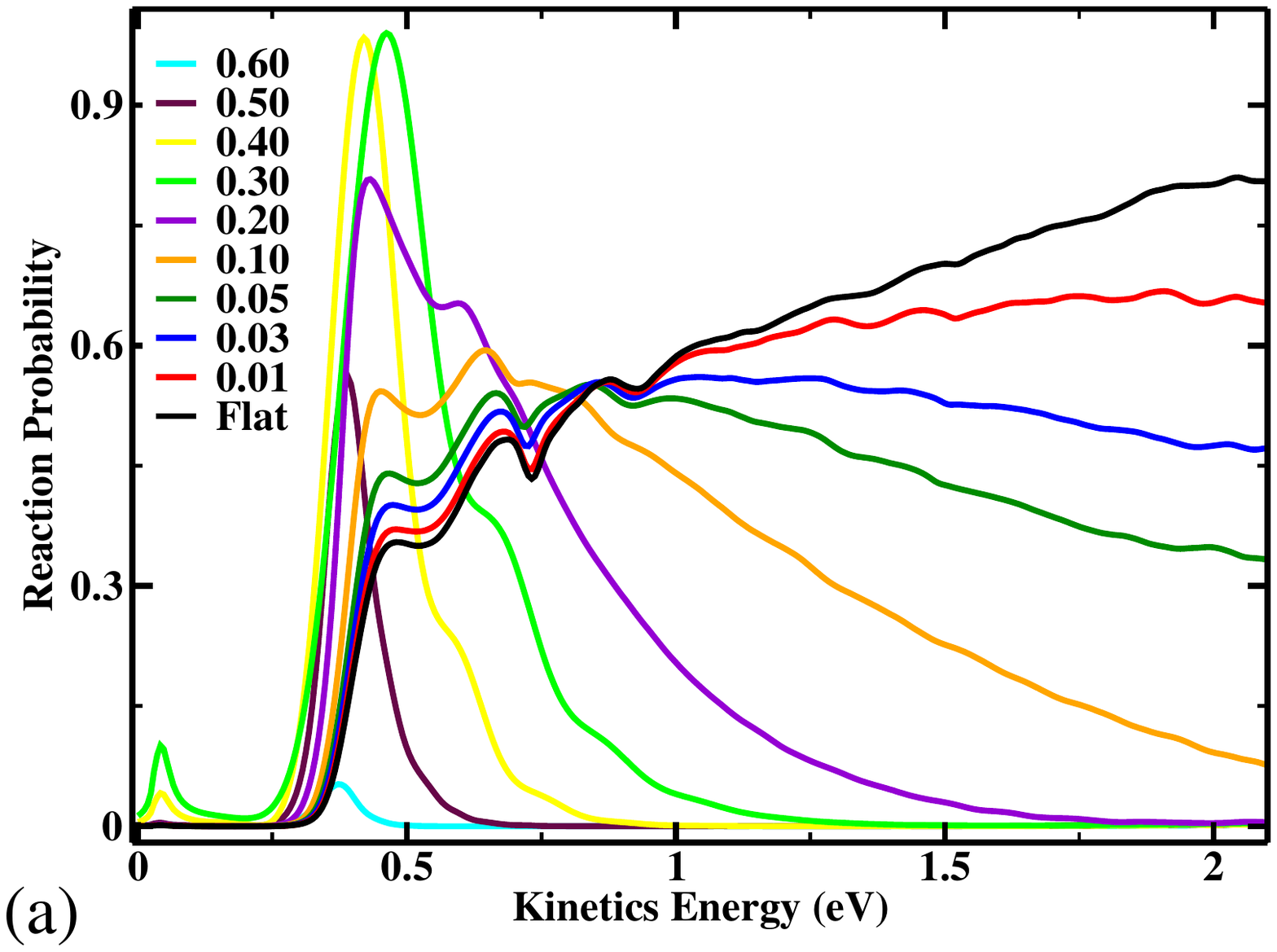}}
     \end{figure}
     
\clearpage
\begin{figure}[h!]
 \centering
  \subfigure[\quad Scattering probability of H$_2$ + H$_2$ in curved spacetime]{
   \includegraphics[width=18cm]{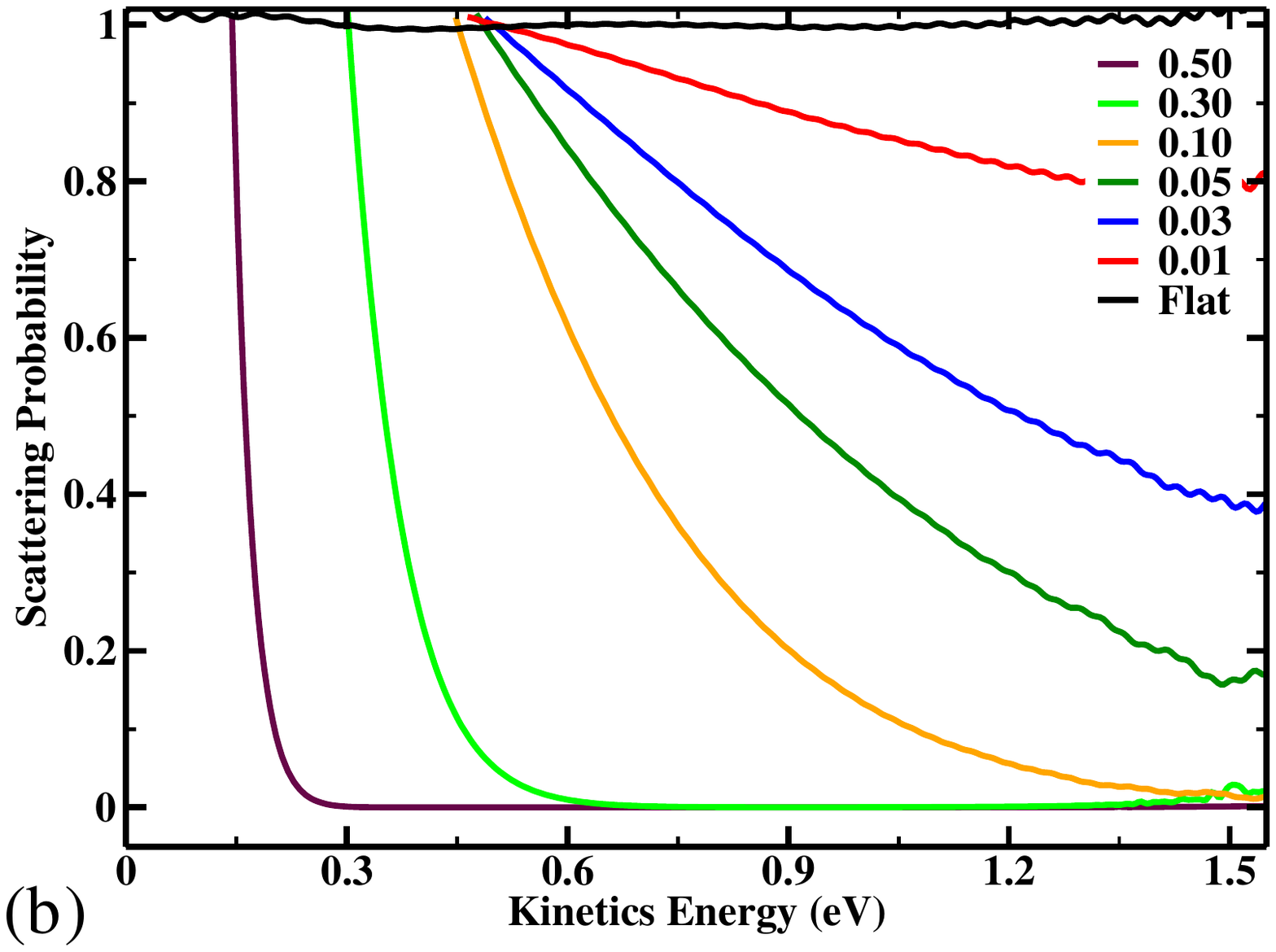}}    
     \end{figure}

\clearpage
\begin{figure}[h!]
 \centering
  \subfigure[\quad Dissociation probability of H$_2$O on Cu(111) in curved spacetime]{
   \includegraphics[width=18cm]{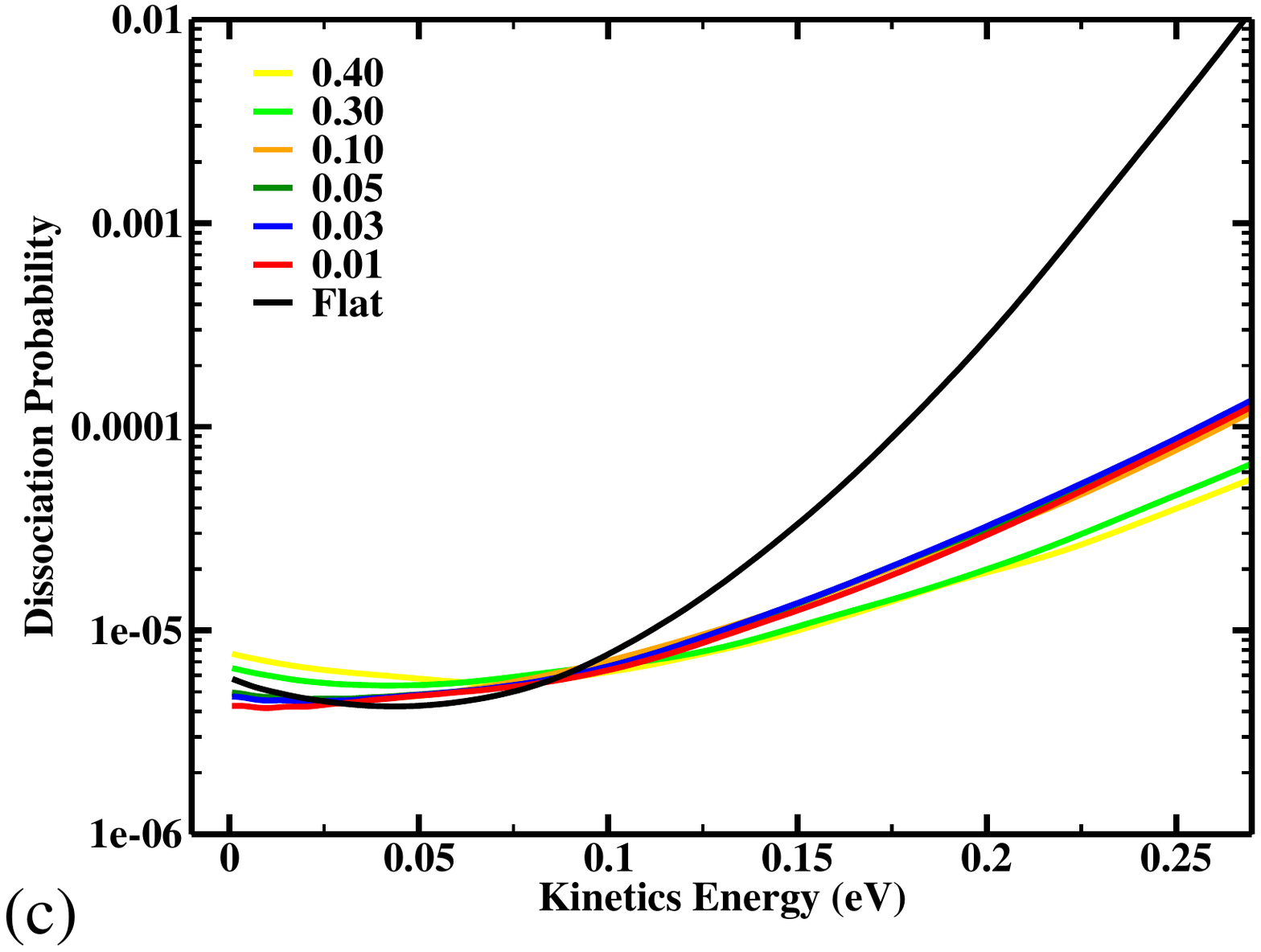}}
     \end{figure}

\clearpage

\clearpage
          
\clearpage
\begin{figure}[h!]
 \centering
  \subfigure[\quad Norm of wave function of H + H$_2$ in curved spacetime]{
   \includegraphics[width=18cm]{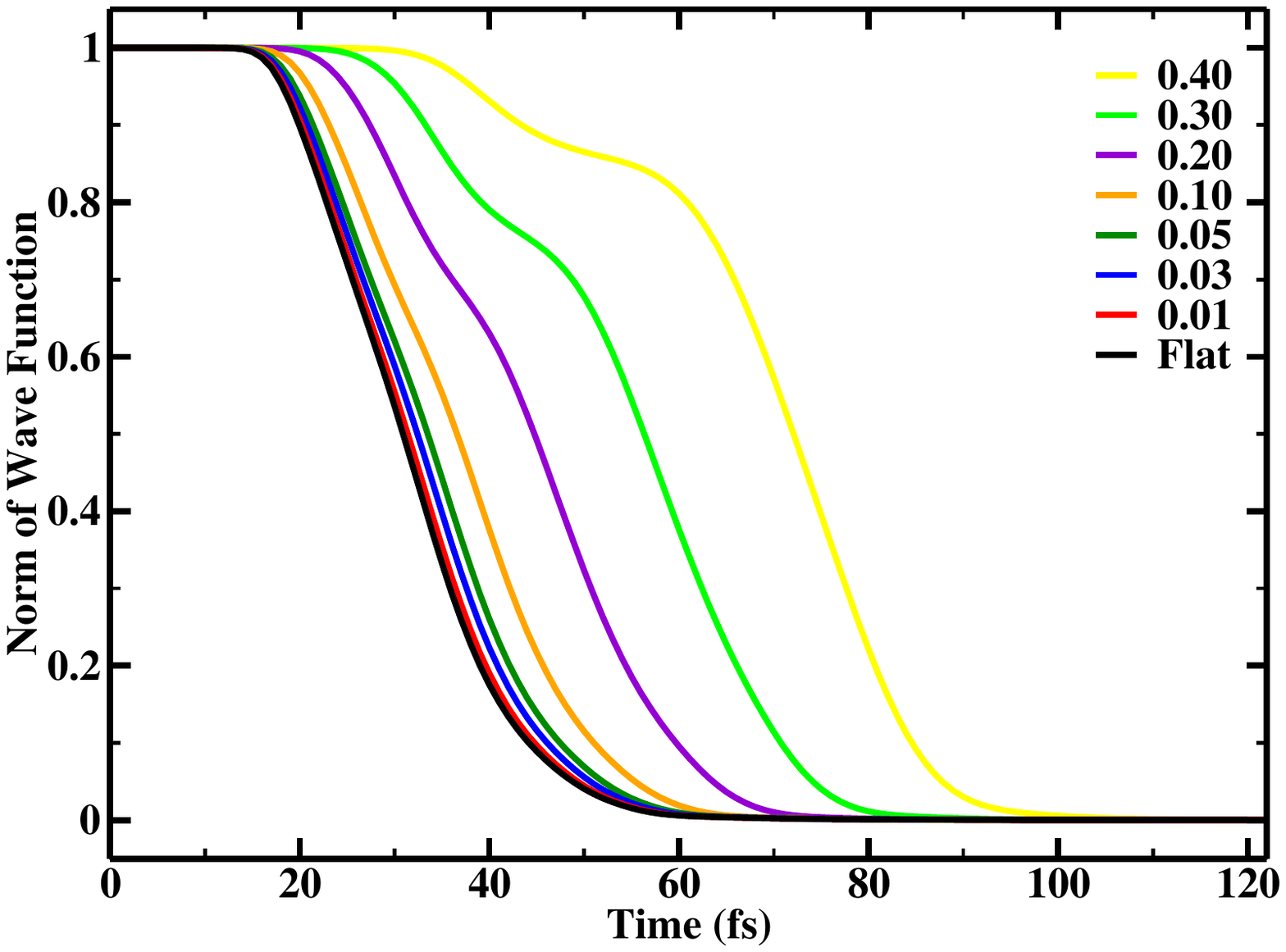}}
    \end{figure}

\clearpage
\begin{figure}[h!]
 \centering
  \subfigure[\quad Norm of wave function of H$_2$ + H$_2$ in curved spacetime]{
   \includegraphics[width=18cm]{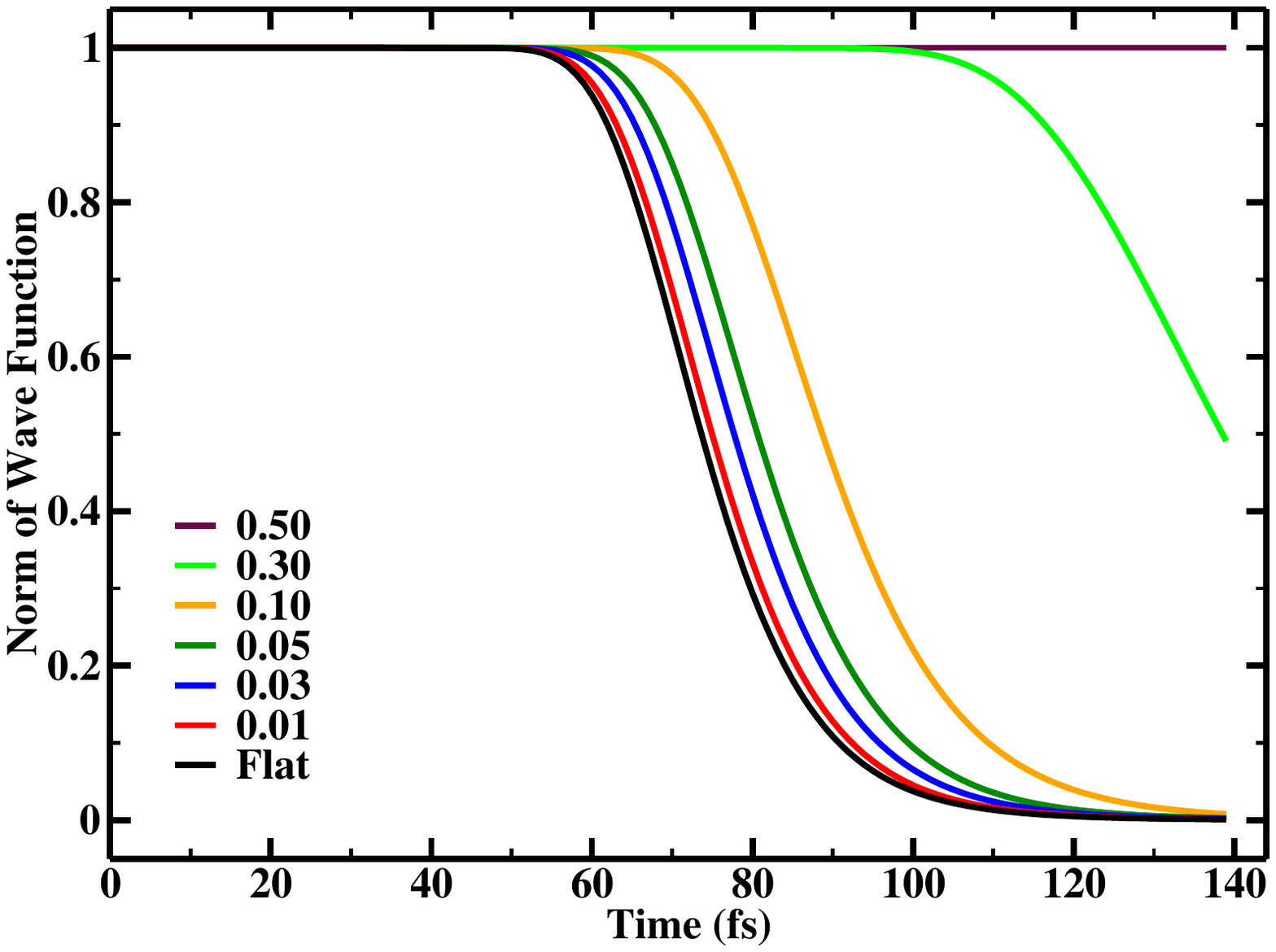}}
    \end{figure}

\clearpage
\begin{figure}[h!]
 \centering
  \subfigure[\quad Norm of wave function of H$_2$O/Cu(111) in curved spacetime]{
   \includegraphics[width=18cm]{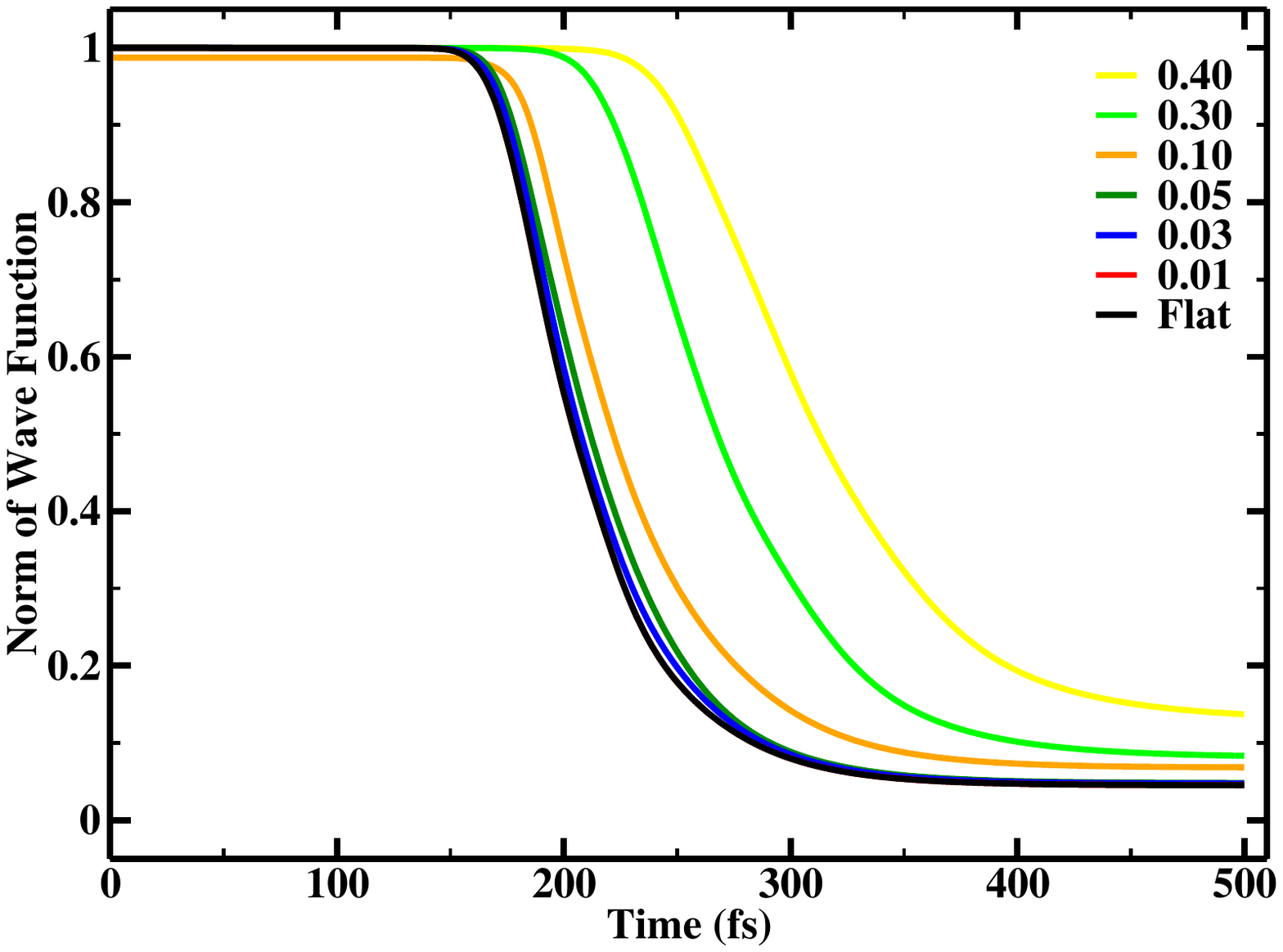}}
    \end{figure}

\clearpage

\clearpage
\begin{figure}[h!]
 \centering
  \subfigure[\quad Expectation of total energy of H + H$_2$ in curved spacetime]{
   \includegraphics[width=18cm]{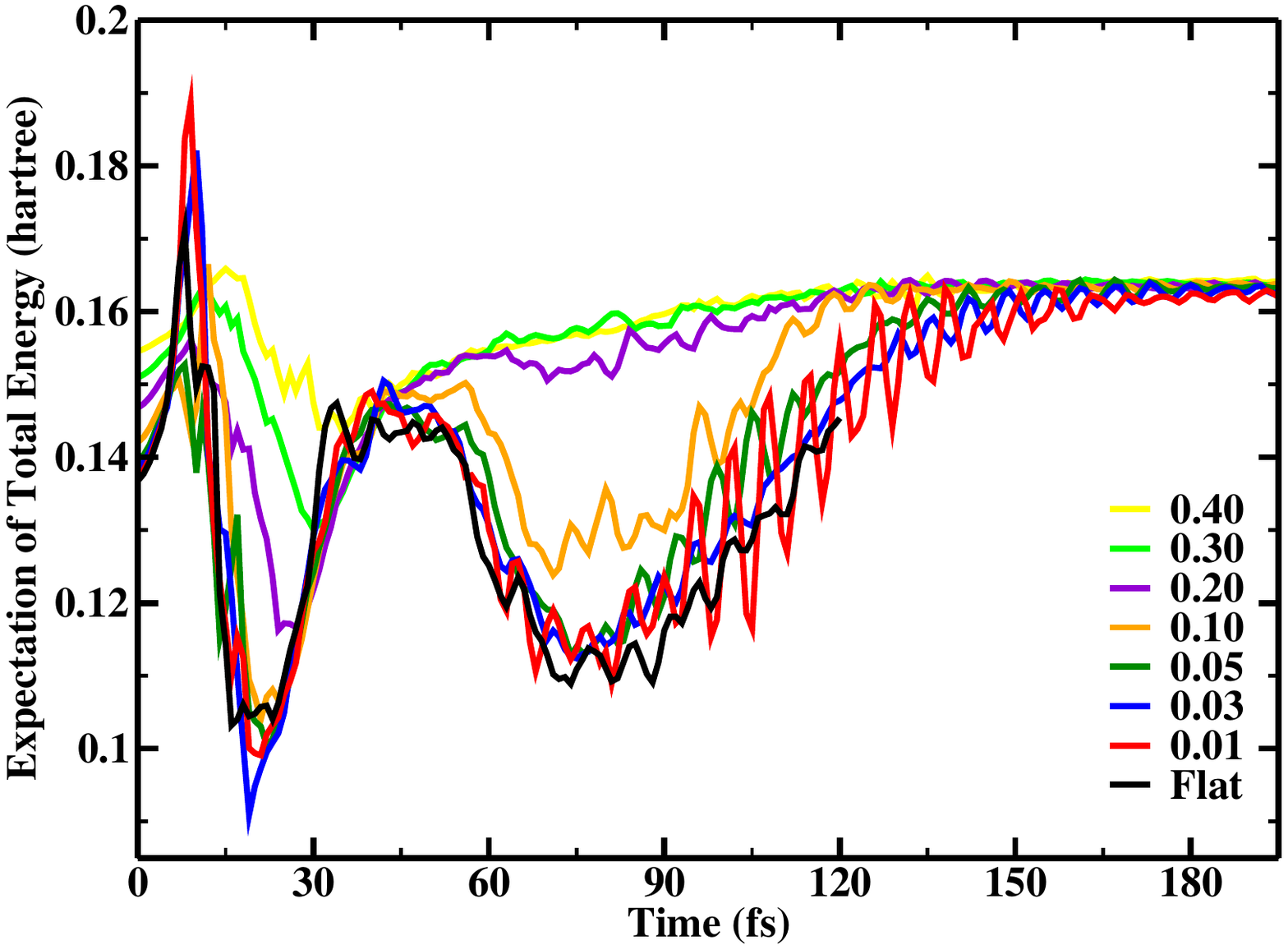}}
     \end{figure}
     
\clearpage
\begin{figure}[h!]
 \centering
  \subfigure[\quad Expectation of total energy of H$_2$ + H$_2$ in curved spacetime]{
   \includegraphics[width=18cm]{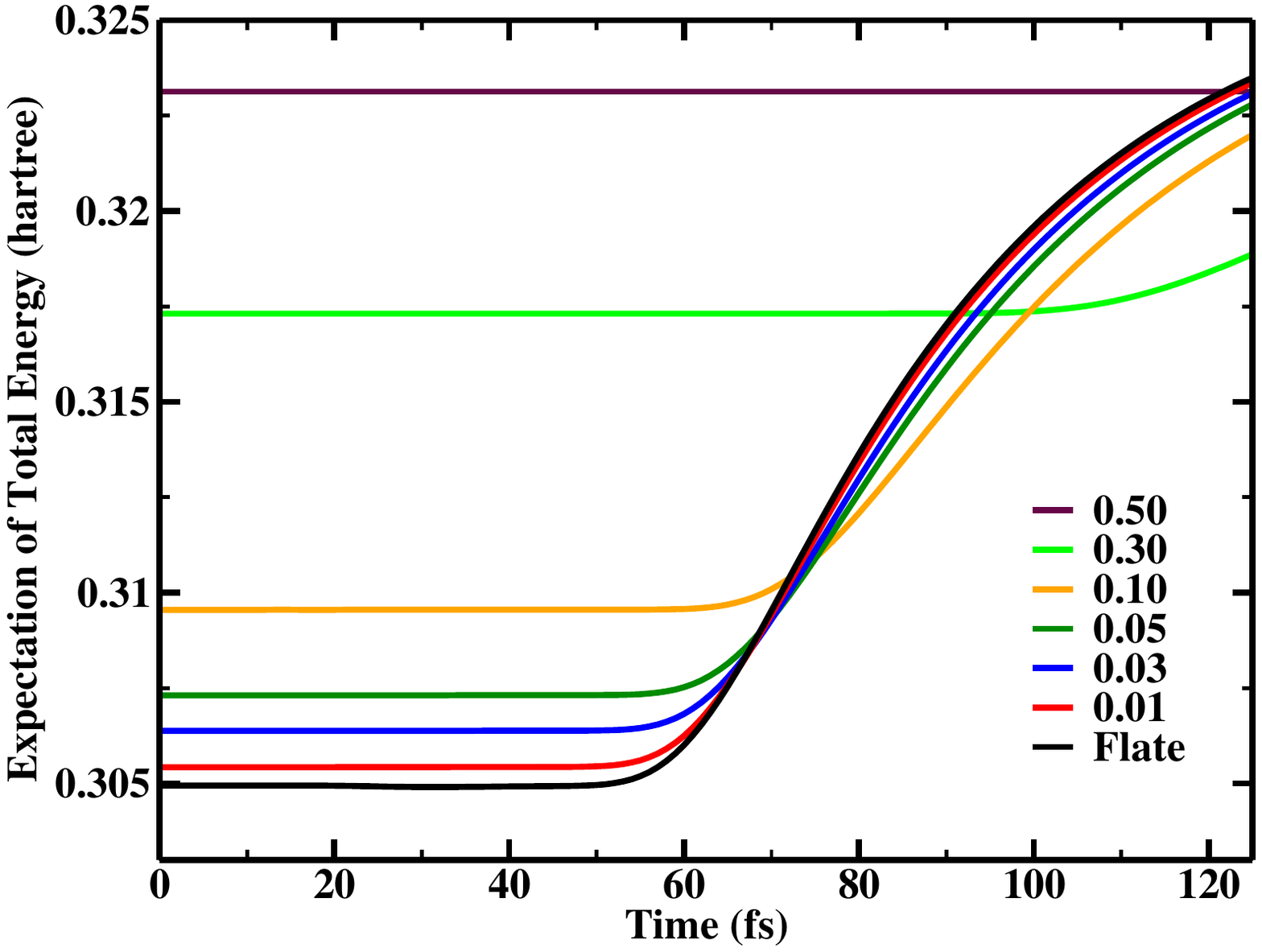}}    
     \end{figure}

\clearpage

\clearpage

\end{document}